%% file: main.tex
\documentclass[letterpaper]{article} 
\usepackage{aaai2026}  
\usepackage{times}  
\usepackage{helvet}  
\usepackage{courier}  
\usepackage[hyphens]{url}  
\usepackage{graphicx} 
\usepackage{natbib}  
\usepackage{caption} 
\usepackage{algorithm}
\usepackage{algorithmic}

\usepackage{url}

\usepackage{newfloat}
\usepackage{listings}
\DeclareCaptionStyle{ruled}{labelfont=normalfont,labelsep=colon,strut=off} 
\floatstyle{ruled}
\newfloat{listing}{tb}{lst}{}
\floatname{listing}{Listing}
\input{utils}
\usepackage{booktabs}
\usepackage{subcaption}
\usepackage{multirow}
\usepackage{pgfplots}
\pgfplotsset{compat=1.18}

\usepackage{xcolor}

\title{Beyond Metadata: Multimodal, Policy-Aware Detection of YouTube Scam Videos}
\author{
    Ummay Kulsum,
    Aafaq Sabir,
    Abhinaya S.B.,
    Anupam Das
}
\affiliations{

    North Carolina State University, Raleigh, NC, USA\\
    ukulsum@ncsu.edu, asabir2@ncsu.edu, asrivid@ncsu.edu, anupam.das@ncsu.edu
}

\usepackage{bibentry}

\begin{document}

\maketitle

\begin{abstract}
\input{sections/abstract}

\end{abstract}

\begin{links}
    {\link{Code}{https://github.com/ummay-kulsum18/VidScamNet}}
    {\link{Dataset}{https://tinyurl.com/VidScam}}
\end{links}

\section{Introduction}
\input{sections/intro}

\section{Related Work}
\input{sections/related_work}

\section{Dataset}
\input{sections/dataset}

\section{Detection of Video-based Scams}
\input{sections/method2}


\section{Results}
\input{sections/result}

\section{Scam Detection in Wild YouTube Videos}
\input{sections/wild_data}

\section{Discussion}
\input{sections/discussion}

\section{Conclusion}
\input{sections/conclusion}

\section*{Acknowledgements}
We thank the anonymous reviewers for their valuable feedback, and Daniel Nolting and Dr. Bradley Reaves (from NC State University) for the YouTube crawler used in our in-the-wild data collection. The opinions, findings, and conclusions expressed here are those of the authors and do not necessarily reflect the views of the participating organizations.
\bibliography{scam}

\appendix
\input{sections/appendix}

\end{document}

%% file: utils.tex
\newcommand{\monetary}{\textit{MonetaryScam}}
\newcommand{\giftcard}{\textit{GiveawayScam}}
\newcommand{\crypto}{\textit{CryptoScam}}

\usepackage{xcolor}
\usepackage{makecell}

\newcommand{\dataset}{\textit{VidScam}}
\newcommand{\model}{\textit{VidScamNet}}

\usepackage{tcolorbox}
\usepackage{enumitem}
\newcommand{\paragraphb}[1]{\vspace{0.05in}\noindent{\bf #1} }

\newcounter{FindingCounter}

\definecolor{block-gray}{gray}{0.85}
\newcounter{rqcounter}
\newcommand{\newrq}[2]{\noindent\refstepcounter{rqcounter}\textbf{RQ\arabic{rqcounter}:} \textit{\textbf{#2}}\label{#1}}

  {\list{}{\leftmargin=0.1in\rightmargin=0.1in}
  \vspace{-4pt}
  \item[]
  }%
  {\endlist
  \vspace{-4pt}
    }
\AtBeginEnvironment{myquote}{\itshape}

\usepackage{pifont}

%% file: sections/abstract.tex
YouTube is a major platform for information and entertainment, but its wide accessibility also makes it attractive for scammers to upload deceptive or malicious content. Prior detection approaches rely largely on textual or statistical metadata, such as titles, descriptions, view counts, or likes, which are effective in many cases but can be evaded through benign-looking text, manipulated statistics, or other obfuscation strategies (e.g., `Leetspeak'), while ignoring visual cues. In this study, we systematically investigate multimodal approaches for detecting YouTube scams. Our dataset consolidates established scam categories and augments them with full-length videos and policy-grounded reasoning annotations. Experiments show that a text-only model using titles and descriptions (fine-tuned BERT) achieves moderate performance (76.61\% F1 score), improving slightly with audio transcripts (77.98\% F1 score). Visual analysis with a fine-tuned LLaVA-Video model performs better (79.61\% F1 score), while a multimodal framework combining titles, descriptions, and video frames achieves the highest performance (82.96\% F1 score). Moreover, the multimodal framework showed greater robustness to adversarial perturbations, with accuracy dropping only 1–3\%, compared to 12–38\% for modality-specific models.
Beyond accuracy, the multimodal framework provides interpretable, policy-grounded reasoning, enhancing transparency and practical utility in automated moderation. Using this approach, we analyzed 6,374 in-the-wild YouTube videos and detected 1,864 scams with explicit reasoning, providing a valuable resource for future research.

%% file: sections/intro.tex
Video has emerged as a dominant medium for disseminating information and engaging audiences on social media~\cite{videoblog}. YouTube, the second most popular social media platform and search engine, exemplifies this trend, with over 500 hours of content uploaded every minute~\cite{youtubestat}. However, its vast reach and accessibility have also made it a hotspot for deceptive and malicious content. Prior research has analyzed diverse scam types on YouTube~\cite{bouma2021first, tripathi2022analyzing, vakilinia2022cryptocurrency, chu2022behind, bouma2025kids}. For example, Bouma-Sims et al.~\cite{bouma2021first} examined giveaway scams offering fake incentives such as gift cards or free premium access, while Tripathi et al.~\cite{tripathi2022analyzing} studied fraudulent apps promising quick financial rewards. These scams often redirect users to external websites or applications to complete surveys, download software, or perform tasks that lead to privacy breaches or malware installation~\cite{bouma2025kids}. More recently, cryptocurrency scams—such as arbitrage bot schemes exploiting flawed smart contracts—have caused substantial financial losses~\cite{li2023towards,vakilinia2022cryptocurrency}. The growing diversity and sophistication of such scams underscore the need for robust and scalable detection mechanisms to safeguard users and preserve the platform’s integrity.

Although prior studies have investigated scams and offered important insights into scam typologies, existing detection approaches largely depend on unimodal cues, particularly textual metadata (e.g., titles, descriptions) or statistical metrics (e.g., view, like, and comment counts). However, as Chu et al.~\cite{chu2022behind} note, these signals can be easily manipulated using fake engagement tactics, rendering them unreliable for scam detection. Moreover, scammers often evade text-based models by crafting benign-looking titles and descriptions that conceal fraudulent intent. In contrast, the visual content of videos often contains richer and more distinctive cues, such as demonstrations of malicious QR codes, fake cryptocurrency dashboards, or misleading giveaway banners, which can serve as stronger indicators of deception. This gap underscores the need for \emph{multi-modal} analysis combining visual and textual signals to improve the robustness and accuracy of YouTube scam detection.
 
With the advent of multimodal large language models (MLLMs) capable of understanding and linking semantics across different modalities, it is timely to explore their potential for detecting video-based scams. Furthermore, existing detection systems primarily produce a likelihood score without grounding their decisions in explicit policy guidelines. However, MLLMs offer the opportunity to incorporate policy grounding, enabling more transparent and interpretable scam detection.

Toward developing a multimodal video-based scam detection system, we aim to address the following research questions:
\newrq{1}{How do video scams manifest across text, audio, and visual modalities?}
This question explores whether scam videos can be reliably detected using only textual metadata (e.g., titles and descriptions) or if audio and visual components provide distinctive and complementary cues. To investigate, we perform a detailed human analysis on a subset of YouTube scam videos, examining modality-specific characteristics and their relevance for automated detection.
\newrq{2}{How can we design a multimodal automated pipeline for video scam detection that incorporates policy-grounded reasoning for greater transparency and interpretability?}
This question highlights the need to combine multimodal understanding with policy-aware reasoning. By jointly analyzing text, audio, and visual cues, the system can capture richer evidence of deception. We fine-tune a MLLM with policy-aligned scam criteria to enable robust cross-modal detection and generate interpretable, policy-consistent explanations.
\newrq{3}{How effective is the multimodal approach in detecting scams from in-the-wild YouTube videos?}
This question assesses the real-world applicability of the proposed approach by evaluating its performance on large-scale, naturally occurring YouTube videos, measuring both detection accuracy and the quality of policy-grounded explanations.
In summary, this study makes the following contributions:
\begin{itemize}[noitemsep,nolistsep,leftmargin=1em]
 \item \textbf{Comprehensive multimodal YouTube scam video dataset:} We curate a new YouTube scam video dataset, \dataset{}, covering monetary, giveaway, and cryptocurrency scams. Unlike prior datasets that include only video IDs and metadata, \dataset{} provides full video content along with policy-aligned reasoning criteria derived from YouTube’s content guidelines.

\item \textbf{Systematic evaluation across modalities:} We perform extensive experiments using text, audio, and visual models to assess modality-specific performance. A text-only BERT model achieves 76.61\% F1 score using titles and descriptions, while a visual-only model (LLaVA-Video-7B) reaches 79.61\% F1 score on video frames—establishing strong baselines for future multimodal research.  

\item \textbf{Multimodal scam detector with interpretable reasoning:} We propose \model{}, a multimodal framework that fuses textual and visual features for improved detection. Beyond classification, \model{} generates interpretable, policy-grounded explanations for its decisions, achieving an F1 score of 82.96\%.

\item \textbf{Large-scale in-the-wild validation:} We evaluate \model{} on 6,374 wild videos collected using scam-related keywords. \model{} identifies 1,864 scam videos and produces YouTube policy–aligned reasoning for each detection, demonstrating its scalability and real-world applicability. We also report the detected scam videos to YouTube.
\end{itemize}


%% file: sections/related_work.tex
\paragraphb{Scams Across Digital Platforms.}
A large body of research has analyzed different types of scams on digital platforms. Most of the work focused on studying a specific scam type such as technical support scam~\cite{gupta2019angel,larson2018using,miramirkhani2016dial,srinivasan2018exposing}, online survey scam~\cite{kharraz2018surveylance}, game hack scam~\cite{badawi2019game}, romance scam~\cite{al2020social,suarez2019automatically}, Pig-Butchering scam~\cite{oak2025hello, burton2024pig}, cryptocurrency scam~\cite{acharya2024conning,liu2024give,li2023towards,li2023double} and SMS scam~\cite{agarwal2025hey,mishra2020smishing}. In addition to this domain-specific analysis, Kolupuri et al.~\cite{kolupuri2025scams} methodologically evaluated the feasibility of Machine Learning (ML) and Deep Learning (DL) approaches for detecting and preventing these fraudulent activities. Kotzias et al.~\cite{kotzias2025ctrl+} studied user exposure, showing that end users most frequently encounter scam domains by following links shared on social media. Our work extends this literature by focusing on scams facilitated through YouTube.

\paragraphb{Video-based Scam/Spam.} 
Prior research has investigated the automated detection of targeted spam and clickbait by analyzing video titles, comments, thumbnails, and metadata~\cite{alberto2015tubespam,chaudhary2013contextual,zannettou2018good}. In a related direction, Chu et al.~\cite{chu2022behind} examined various tactics used by creators to monetize content through deceptive means. Llavendhan et al.~\cite{ilavendhan2024optimizing} focused on detecting spam within YouTube comments and do not analyze the video content itself, making their work fundamentally different from ours.
Several studies have explicitly focused on detecting YouTube scam videos. Bouma-Sims et al.~\cite{bouma2021first} conducted an exploratory analysis using organic searches and found that basic metadata, such as video age, view count, and channel size, alone were insufficient for reliably identifying scams. Tripathi et al.~\cite{tripathi2022analyzing} compared textual features, like titles and descriptions, with statistical metadata (e.g., view count, likes, video length, comments) and found that text-based features were more effective for detecting monetary scams. Similarly, Li et al.~\cite{li2023towards} demonstrated that classifiers trained on textual metadata could successfully detect cryptocurrency arbitrage bot scams, highlighting the predictive value of textual information in this domain.

\paragraphb{Distinction from Prior Work.}
Our study differs from prior efforts in three key ways. First, unlike existing approaches that rely primarily on textual or statistical metadata, we incorporate the video modality, enabling multimodal detection of scam videos on YouTube, which is more resistant to evasion techniques. Second, we focus on multiple scam categories to improve the generalizability of our detection approach, rather than limiting it to a single scam type. Finally, while these traditional models lack explainability, our approach integrates scam reasoning criteria grounded in YouTube’s official policy on ``Spam, deceptive practices, and scams”~\cite{youtube_policy}, providing a level of explainability that is essential for effective platform governance.

%% file: sections/dataset.tex
\label{sec:dataset}
We utilize three publicly available YouTube scam video datasets from prior studies, each focusing on a distinct scam category: monetary~\cite{bouma2021first}, giveaway~\cite{tripathi2022analyzing}, and cryptocurrency scams~\cite{li2023towards}. These datasets were chosen for their open accessibility and coverage of diverse scam types. Each dataset provides YouTube video IDs, titles, descriptions, and associated metadata. The corresponding videos were retrieved using the Python library \texttt{yt-dlp}~\cite{yt-dlpgit} by providing the video IDs. We name this consolidated dataset as \dataset{}.

\input{tables/criteria}
\input{figures/examples}
\subsection{Adapting Existing Datasets} 
Bouma-Sims et al. \cite{bouma2021first} compiled a dataset of 3,700 YouTube videos related to various scams, including gift card scams, game currency scams, tech support scams, and bank support scams. The authors manually reviewed and labeled 668 videos as scams, while the remaining videos were classified as non-scams. However, we were able to download only 2,071 videos, of which 146 were labeled scams. Throughout the paper, we refer to this dataset as \giftcard{}.

Tripathi et al.~\cite{tripathi2022analyzing} focused on content promoting financial fraud, such as cash gift offers, get-rich-quick schemes, and pyramid schemes, which they categorized as monetary scam videos. Their dataset consists of 1,292 manually annotated YouTube videos, including 419 monetary scam videos, 873 non-scam videos. We successfully downloaded 1,188, including 278 labeled scams. Throughout the paper, we refer to this dataset as \monetary{}.

Li et al. \cite{li2023towards} investigated cryptocurrency arbitrage bot scams and developed a classification model called CryptoScamHunter, which detects such fraudulent content on YouTube. Their model was trained using 2,000 ground-truth YouTube video titles and descriptions. However, we could not obtain the ground-truth videos, as the video IDs were unavailable. Nonetheless, we accessed a set of video IDs classified as scams by CryptoScamHunter and downloaded 580 of these videos. Throughout the paper, we refer to this dataset as \crypto{}.

\paragraphb{Ethical Considerations.} The datasets consist exclusively of publicly available YouTube videos and do not include user-level personal information. Any incidental personal identifiers were filtered or anonymized prior to use.
\vspace{-5.5pt}
\subsection{Annotation and Dataset Validation} 
\label{subsec:human_annotation}
The three existing YouTube scam datasets primarily focus on metadata (titles, descriptions, and identifiers) but have two key limitations: (i) they lack explicit annotations of policy violations that define scam content, and (ii) they do not attribute these violations to specific modalities or video properties such as the title, description, or visual content. To address this limitation, we re-annotated a subset of the dataset with two additional labels, i.e., scam modality and scam criteria. 
The modality label identifies where scam cues appear  such as text, audio, and visual and is annotated using a binary scheme, where each modality is independently marked as present or absent (allowing multiple modalities per video). In contrast, the criteria labels are grounded in YouTube’s policies on spam, deceptive practices, and scams~\cite{youtube_policy}.
Following Bouma-Sims et al.~\cite{bouma2021first}, we extract and consolidate scam-related criteria defined in YouTube’s content policies.
We present our final set of scam criteria with representative examples in Table~\ref{tab:scam_criteria}, and illustrate instances from the dataset in Figure~\ref{fig:criteria_examples}. Collectively, these criteria capture the manipulative tactics designed to deceive users and redirect them into fraudulent ecosystems, while our modality annotations reveal how such tactics are operationalized across text, audio, and visual modalities.

During annotation, we observed that many “scam” videos did not result in direct financial loss. To capture this nuance, we introduced an additional labeling dimension 
for scam implication with three categories: (i) in-game financial or asset gain, (ii) real-world financial or material gain, and (iii) redirection to external websites or apps that may host malware or collect personal data. These labels offer finer granularity in characterizing scam tactics, highlighting how promises of financial or in-game rewards are often used to lure users into fraudulent ecosystems.

\paragraphb{Annotation Process.} 
\label{sec:human_annotation}
With a fixed set of scam criteria, we analyze 200 randomly selected videos from the consolidated dataset. Among these videos, 120 were labeled as scams. For the annotation process, three researchers with 3-5 years of experience in online safety research annotated videos in the dataset in batches of 10-20. For each video, annotators applied scam criteria, scam implication, modality, and a label indicating ``scam", ``non-scam". Each video was reviewed in full, including audio, title, description, and visual content. After every batch, they convened and discussed their application of the criteria to reach a consensus on their mental models. In some cases where the legitimacy of external apps/websites mentioned in the video was unclear, annotators verified with two widely used URL checkers, NordVPN~\cite{NordVPN} and VirusTotal~\cite{VirusTotal}, labeling the video as scam/non-scam. Moreover, for non-english videos, annotators used YouTube's auto-captioning and auto-translation features.

\paragraphb{Inter-rater reliability.} To assess inter-rater reliability, we computed \emph{Krippendorff’s alpha}~\cite{mcdonald2019reliability, krippendorff2011computing} after each batch, using a multi-label implementation with the Dice similarity metric~\cite{MtrevisoKrippendorffalphaPython}. Krippendorff’s alpha ($\alpha$) is well-suited for multi-label categorical annotation with multiple raters. Iterative annotation continued until substantial agreement ($\alpha > 0.8$) was achieved after nine iterations, covering 115 videos. The remaining 85 videos were then distributed among annotators for independent labeling. Detailed reliability scores in all iterations are reported in Appendix~\ref{Appendix:1}. For quality control, the first two batches were structured by scam type (all scam, all non-scam), while later batches were randomized across classes. For non-scam videos, annotators additionally provided a short descriptive summary confirming the absence of deceptive practices.

\begin{figure}[!t]
\centering
\includegraphics[width=0.95\columnwidth]{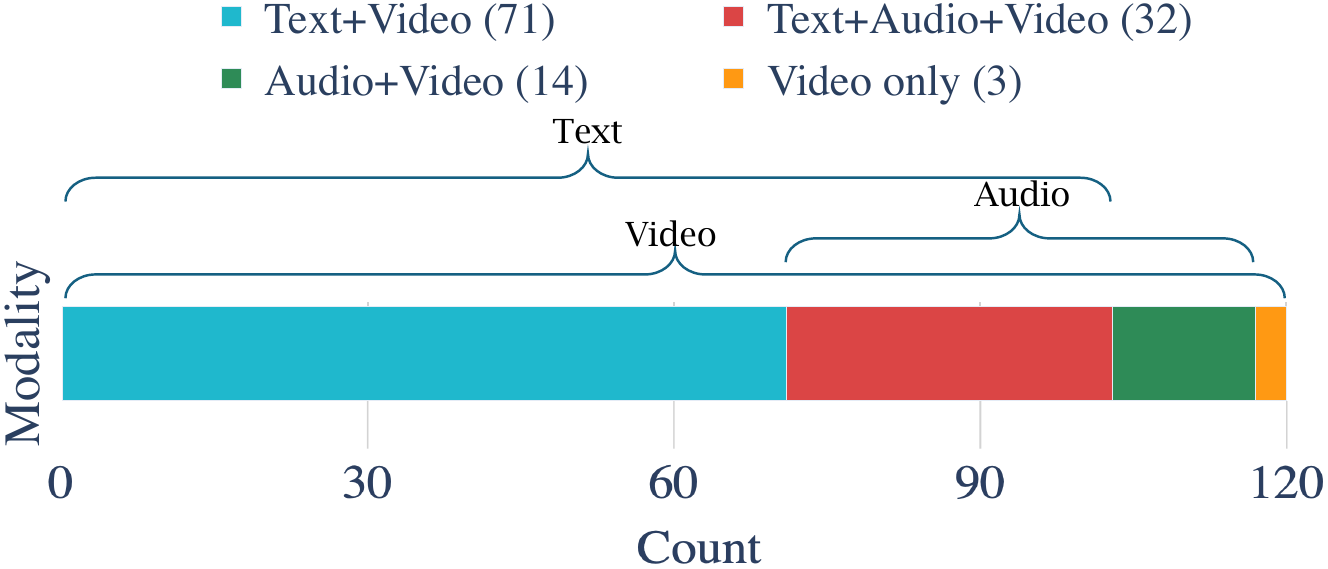}
\caption{Scam cues across modality from annotated dataset.} 
\label{fig:modality}
\end{figure}

Figure~\ref{fig:modality} presents the distribution of combinations of modality in the annotated scam videos. 
The video title and description are grouped as the text modality. The majority of the scam videos relied on multimodal cues rather than a single modality. Specifically, 32 videos contained indicators across all three modalities (text, audio, and video), while 71 combined textual metadata with visual cues (text+video), and 14 leveraged audio and video (audio+video). Only three videos relied exclusively on visual cues, and no videos were annotated as text-only, audio-only, or text+audio.
This underscores that multimodal information provides richer and more reliable indicators for identifying scams. Figure~\ref{fig:titleVideo} illustrates a typical [text+video] case: the title advertises a “hack” for obtaining free in-game currency, while the video frames explicitly display a malicious link to claim the reward. By contrast, Figure~\ref{fig:video} shows a [video-only] example, where the title and description appear benign (e.g., referencing “FrontRunning Bot BSC”), yet the video frames reveal strong scam cues, such as invitations to join private Telegram channels to purchase cryptocurrency bots. 
Moreover, most scam videos contained only background music or no audio at all. We also observed recurring visual patterns that serve as strong scam indicators: on-screen URLs or QR codes linking to external websites, textual instructions embedded in the video such as “Browse this website,” “Click on Generate,” or “Link in the Description,” and direct contact information for private Telegram, Discord, or WhatsApp groups. Many scam videos simulate “human verification” steps, requiring users to complete surveys, provide personal information, or install third-party applications to claim promised rewards. Some even encourage risky behavior, such as disabling antivirus software to install fraudulent tools. Collectively, these results indicate that while textual metadata is informative, combining visual cues with text provides the most reliable indicators of scams, underscoring the value of multimodal analysis.

\input{tables/criteria_dist}
\input{tables/criteria_narrow_dist}
Table~\ref{tab:scam_distribution} summarizes the prevalence of scam criteria in the annotated datasets. The most frequently observed categories include: `Sends audiences to sites that can spread malware, try to gather personal information or other sites that have a negative impact', `Offers cash gifts, get rich quick schemes, or pyramid schemes', and `Gets clicks, views, or traffic off YouTube by promising viewers that they'll make money fast'. Moreover, Table~\ref{tab:scam_narrow_distribution} presents the annotation result for scam implication. Here, financial or material gain emerged as the most prominent one, appearing in 73 times, and in game financial or asset gain appears 30 times. Together, these distributions highlight scammers’ reliance on monetary and asset-based incentives as primary lures.

\begin{tcolorbox}[colback=lightgray,colframe=lightgray, boxsep=1pt,left=2pt,right=2pt,top=0pt,bottom=0pt]
\textbf{Takeaway:} Majority of the scam videos leave distinctive scam cues in both the textual and visual modality. 
\end{tcolorbox}

\input{figures/modality}

For this study, we initially treated the labels from the existing datasets as ground truth. However, after annotation, the researchers disagreed with the ground truth labels for 13 videos, all of which had been originally labeled as scams. To further investigate potential mislabeling, annotators reviewed all scam videos in the merged dataset. As a result, 9 videos from the \monetary{} dataset were reclassified as non-scams, as they discussed legitimate income-generating activities such as freelancing, affiliate marketing, and blogging. Other examples included informational videos about Google Pay or recordings of award ceremonies. Similarly, 5 videos from the \giftcard{} dataset were relabeled as non-scams, which include content such as discussions about why certain Clash of Clans game hacks are scams or music videos unrelated to fraudulent activities. Finally, since \crypto{} videos were labeled using the automated tool CryptoScamHunter, a substantial number of false positives were identified. 41 videos from this dataset were reclassified as non-scams, including cryptocurrency tutorials, podcasts, and unrelated general content such as cooking videos and fishing vlogs. For the final dataset, all identified videos were relabeled as non-scams to ensure higher data quality (mislabeling rate is in Appendix~\ref{Appendix:data_quality}).

\input{tables/data_split}

\subsection{Training and Test Sets}
Using the consolidated dataset, we constructed the training and test sets for our experiments. The training set has a 1:2 ratio of scam to non-scam videos, comprising a total of 500 scam videos: 200 from \monetary{}, 100 from \giftcard{}, and 200 from \crypto{}. From the combined pool of non-scam content across all three datasets, 1,000 non-scam videos were randomly sampled. The test set comprises the remaining videos and does not adhere to a fixed ratio of scams to non-scams. A detailed breakdown of the training and test sets is provided in Table~\ref{tab:data_split}.

%% file: tables/criteria.tex
\begin{table*}[!t]
\centering

\footnotesize
\begin{tabular}{p{0.08\linewidth} p{0.31\linewidth} p{0.52\linewidth}}
\toprule
\textbf{Criteria} & \textbf{Description} & \textbf{Example} \\
\midrule
Criteria 1 & Claims to commit a crime on behalf of the user, regardless of whether it actually does. & Video instructing viewers to manipulate a company’s customer service to obtain free goods or sharing content that directly violates platform policies. \\
\hline
Criteria 2 & Purports to provide an unbounded giveaway that offers unlimited free items without rules, limit or end. & Claiming to generate unlimited gift card codes or infinite in-game currency through third-party applications. \\
\hline
Criteria 3 & Promises viewers they'll see something but instead directs them off-site. & Video that promises a movie clip but instead links to an external streaming site. \\
\hline
Criteria 4 & Gets clicks, views, or traffic off YouTube by promising viewers that they'll make money fast. & Video that advertises rapid financial or in-game gains in order to redirect users to malicious applications or external websites(e.g., ``Get a free \$200 Nike Gift Card'', ``Make \$50 per day using Crypto arbitrage bot'')\\
\hline
Criteria 5 & Sends audiences to sites that can spread malware, try to gather personal information or other sites that have a negative impact. & Video that explicitly instructs viewers to follow links in the description or visit external sites that are potentially harmful. \\
\hline
Criteria 6 & Offers cash gifts, get rich quick schemes, or pyramid schemes. & Video that promotes unrealistic promises of financial or in-game rewards, such as fraudulent mobile game hacks or earning money by playing games.\\
\hline
Criteria 7 & Impersonates an individual, company, or organization. & Video that advertises fake customer support number posing as Amazon support. \\
\bottomrule
\end{tabular}
\caption{Scam criteria used for annotation of YouTube videos, along with representative examples.}
\label{tab:scam_criteria}
\end{table*}

%% file: figures/examples.tex
\begin{figure*}[t]
    \centering
    \begin{subfigure}[b]{0.3\textwidth}
        \centering
        \includegraphics[width=\textwidth]{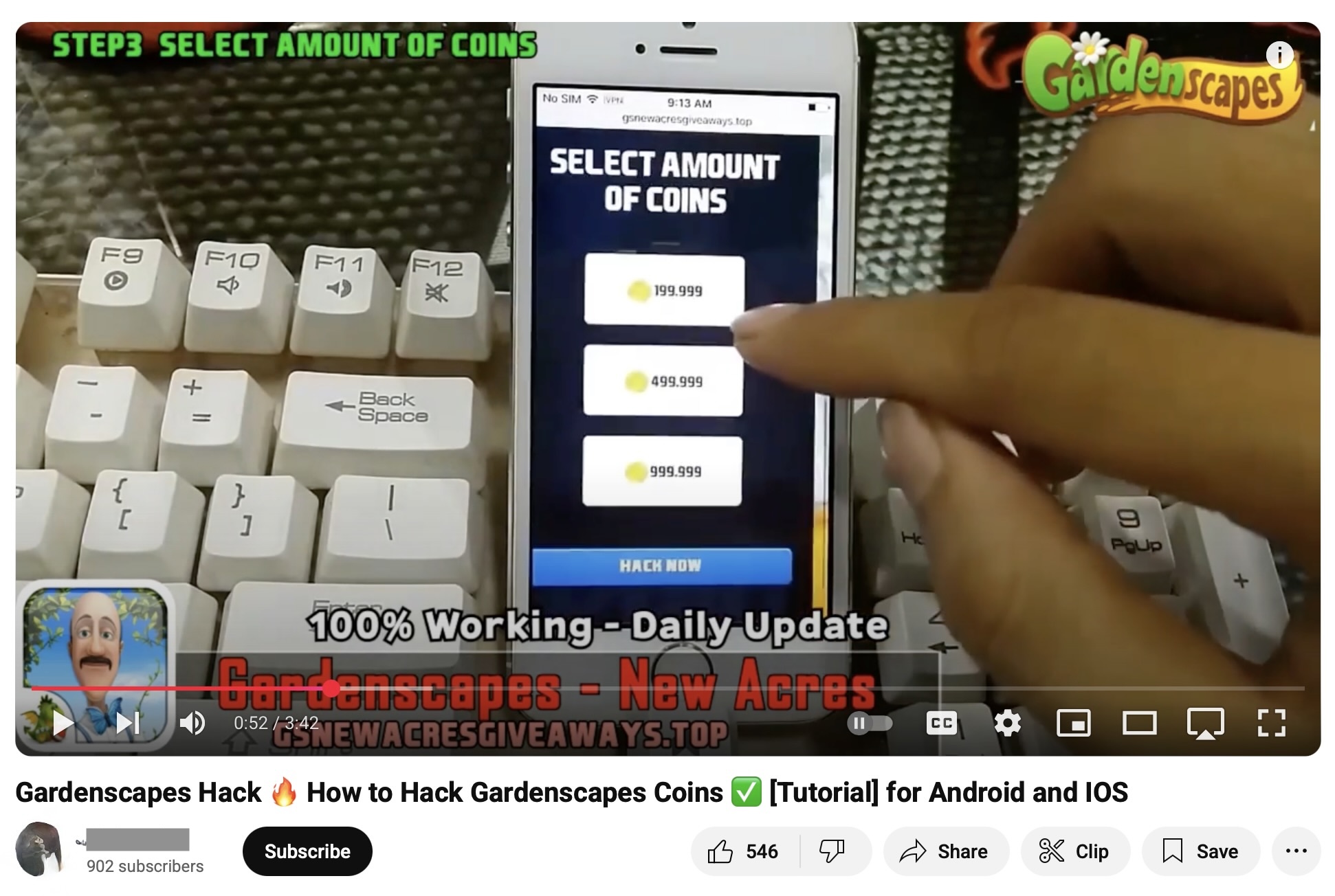}
        \caption{}
        \label{fig:sub1}
    \end{subfigure}
    \hfill
    \begin{subfigure}[b]{0.3\textwidth}
        \centering
        \includegraphics[width=\textwidth]{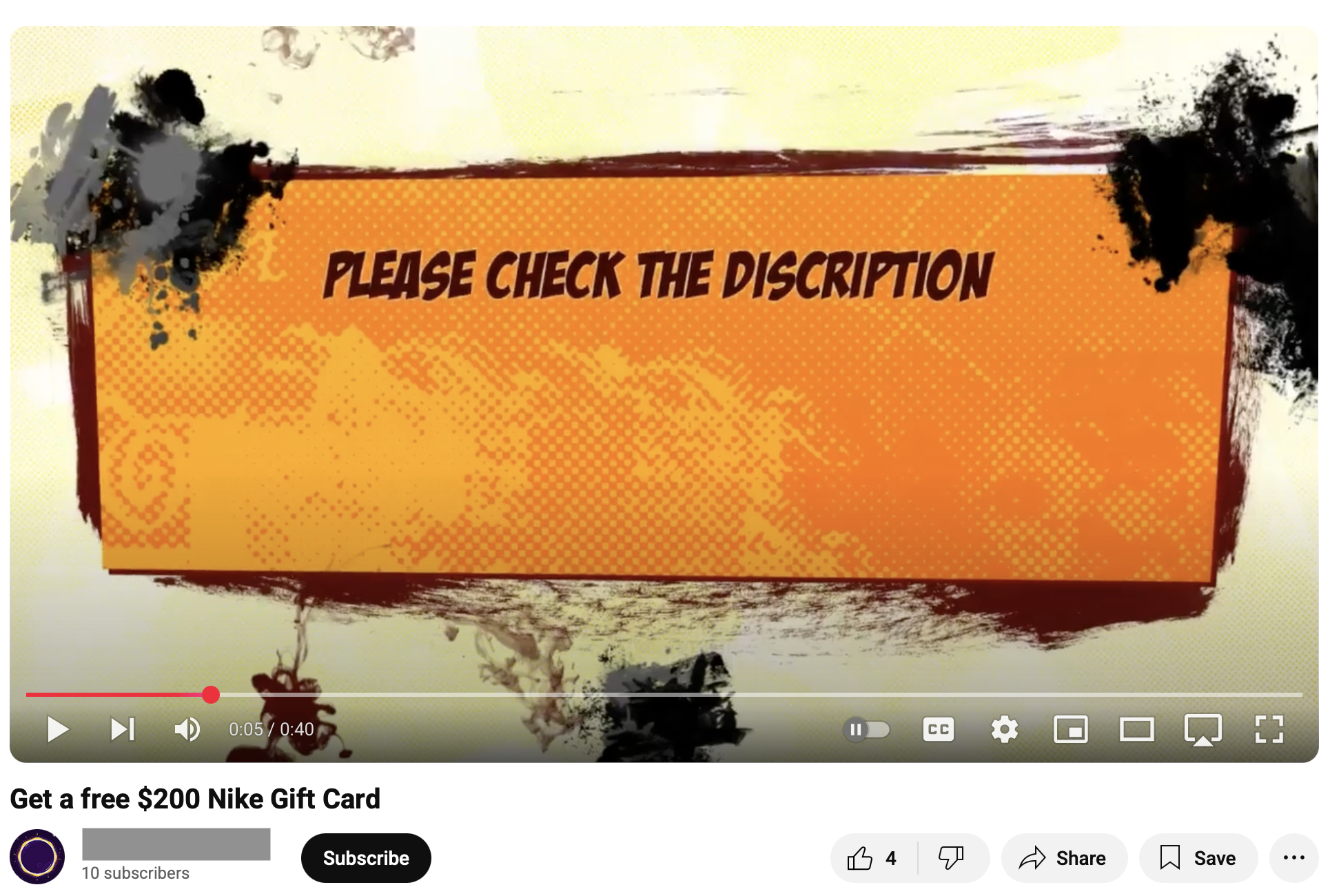}
        \caption{}
        \label{fig:sub2}
    \end{subfigure}
    \hfill
    \begin{subfigure}[b]{0.3\textwidth}
        \centering
        \includegraphics[width=\textwidth]{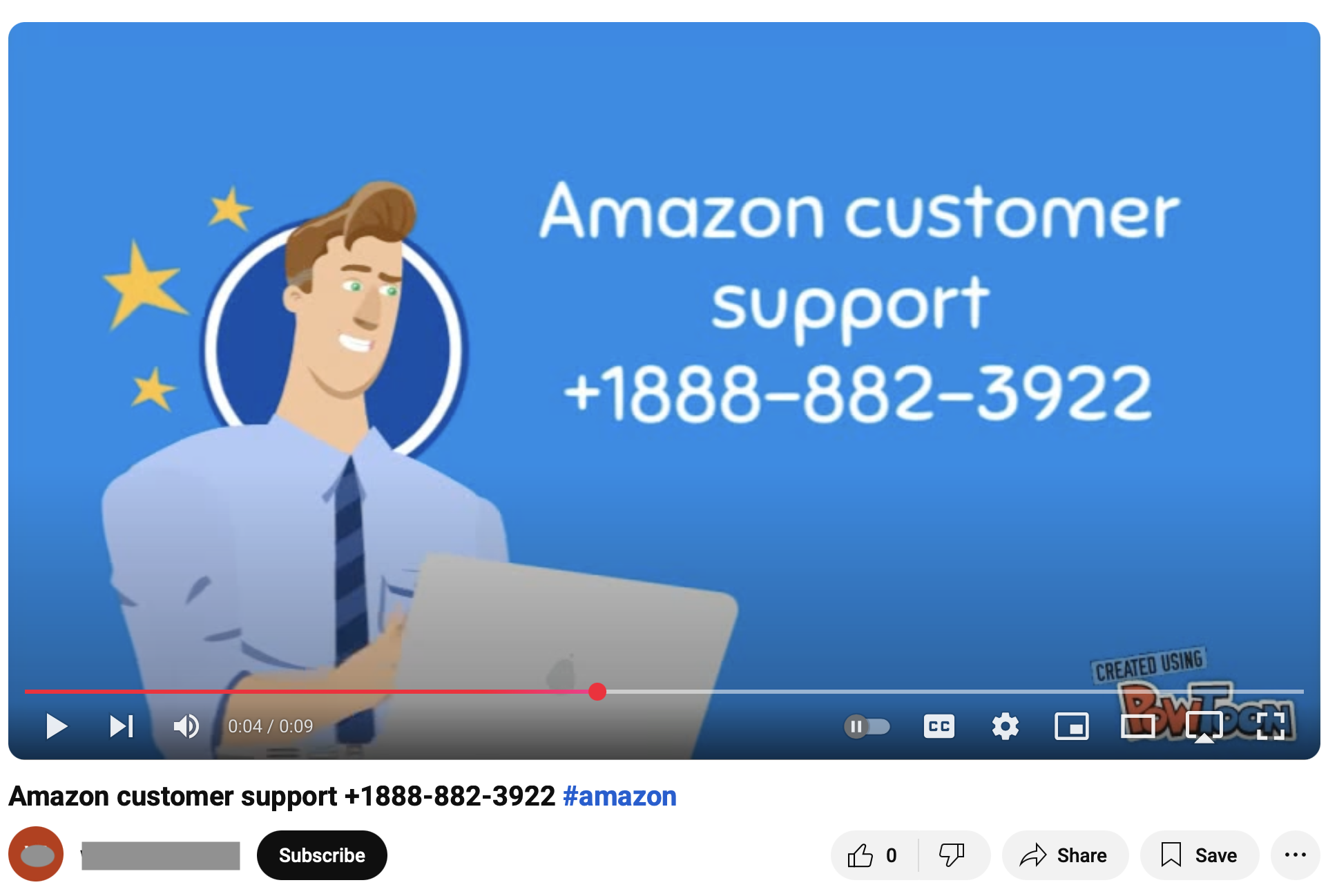}
        \caption{}
        \label{fig:sub3}
    \end{subfigure}
    \caption{Example scam videos. (a, b) illustrate scams that redirect users to external websites, promote “get rich quick” schemes, and attempt to drive clicks, views, or traffic off YouTube. (c) impersonates Amazon with a likely fraudulent contact number.}
    \label{fig:criteria_examples}
\end{figure*}

%% file: tables/criteria_dist.tex
\begin{table}[!t]
\centering

\resizebox{\linewidth}{!}{
\begin{tabular}{cccccccc}
\toprule
Crit. 1 & Crit. 2 & Crit. 3 & Crit. 4 & Crit. 5 & Crit. 6 & Crit. 7 \\
\toprule
3 & 12 & 0 & 88 & 105 & 102 & 16 \\

\bottomrule
\end{tabular}
}

\caption{Scam criteria distribution in annotated dataset.}
\label{tab:scam_distribution}
\end{table}

%% file: tables/criteria_narrow_dist.tex
\begin{table}[!t]
\centering

\resizebox{\linewidth}{!}{
\begin{tabular}{p{0.85\linewidth}r}
\toprule
\textbf{Scam Criterion} & \textbf{Count} \\
\midrule
Financial or material gain & 73 \\
\hline
In game financial or asset gain & 30 \\
\hline
Redirect to website or app that can be malware or collect personal information & 105 \\
\bottomrule
\end{tabular}
}

\caption{Distribution of scam type in annotated dataset.}
\label{tab:scam_narrow_distribution}
\end{table}

%% file: figures/modality.tex
\begin{figure*}[!t]
    \centering
    \begin{subfigure}[t]{0.45\textwidth}
        \centering
        \includegraphics[width=\linewidth]{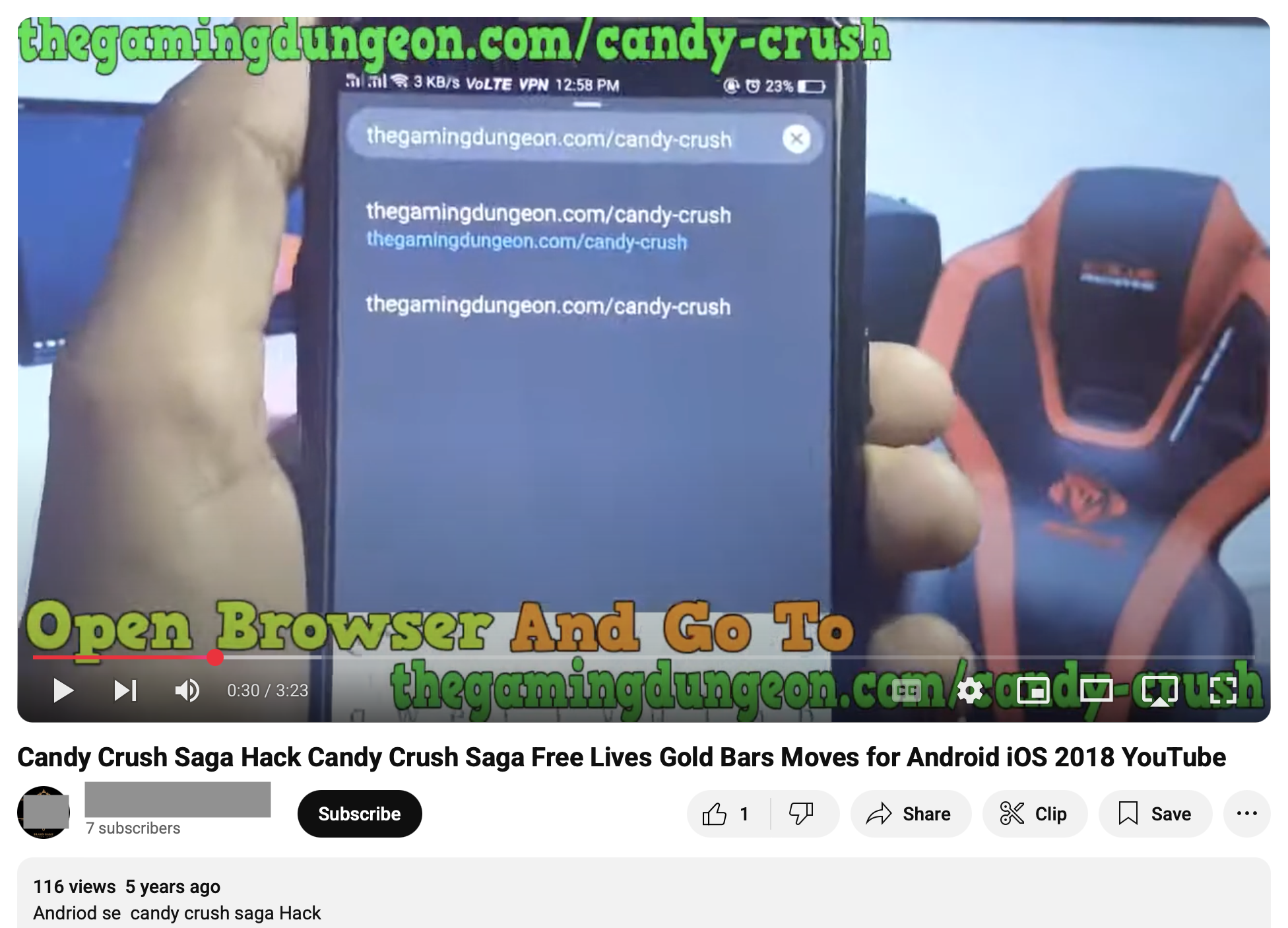}
        \caption{}
        \label{fig:titleVideo}
    \end{subfigure}
    \hfill
    \begin{subfigure}[t]{0.45\textwidth}
        \centering
        \includegraphics[width=\linewidth]{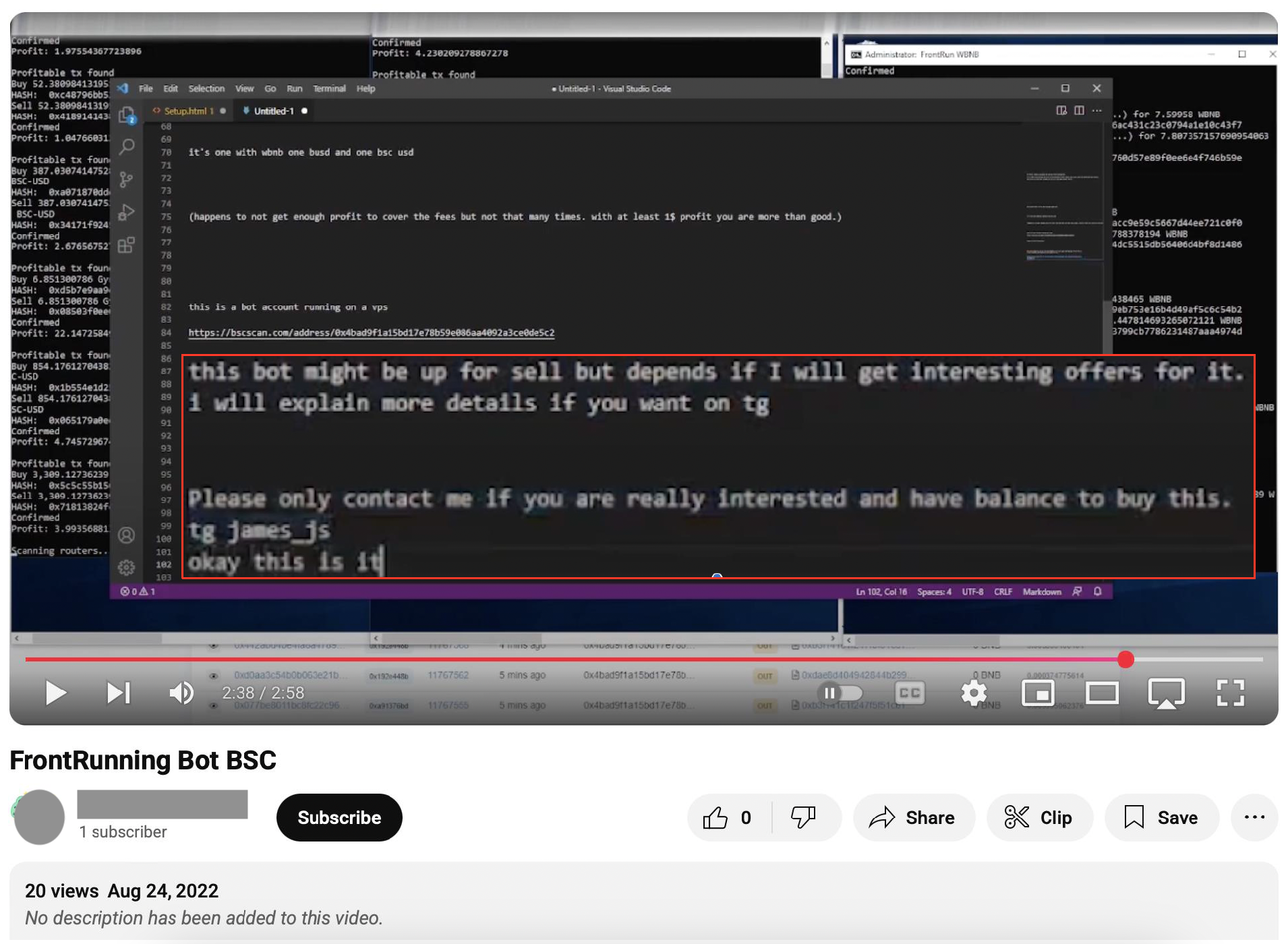}
        \caption{}
        \label{fig:video}
    \end{subfigure}
    \caption{Modality of scam videos. (a) combines text and visuals: a title advertising a game hack with frames showing instructions and a malicious link; (b) relies on visual deception, directing viewers to a Telegram channel for a crypto arbitrage bot.}
    \label{fig:twofigs}
\end{figure*}

%% file: tables/data_split.tex
\begin{table}[!t]
\centering

\small
\resizebox{0.9\columnwidth}{!}{
\begin{tabular}{lcccc}
\toprule
Label             & Dataset        & Train & Test & Total \\ \midrule
\multirow{4}{*}{Scam} & \monetary{}       & 200   & 69   & 269   \\
                      & \giftcard{}       & 100   & 41   & 141   \\
                      & \crypto{}         & 200   & 339  & 539   \\ \cline{2-5} 
                      & \dataset{} & 500   & 449  & 949   \\ \hline
Non Scam              & \dataset{} & 1000  & 1890 & 2890  \\ \hline
All                   & \dataset{} & 1500  & 2339 & 3839  \\ \bottomrule
\end{tabular}}
\caption{Train and Test Dataset Distribution}
\label{tab:data_split}
\end{table}

%% file: sections/method2.tex
\begin{figure}[!t]
\centering
\includegraphics[width=\linewidth]{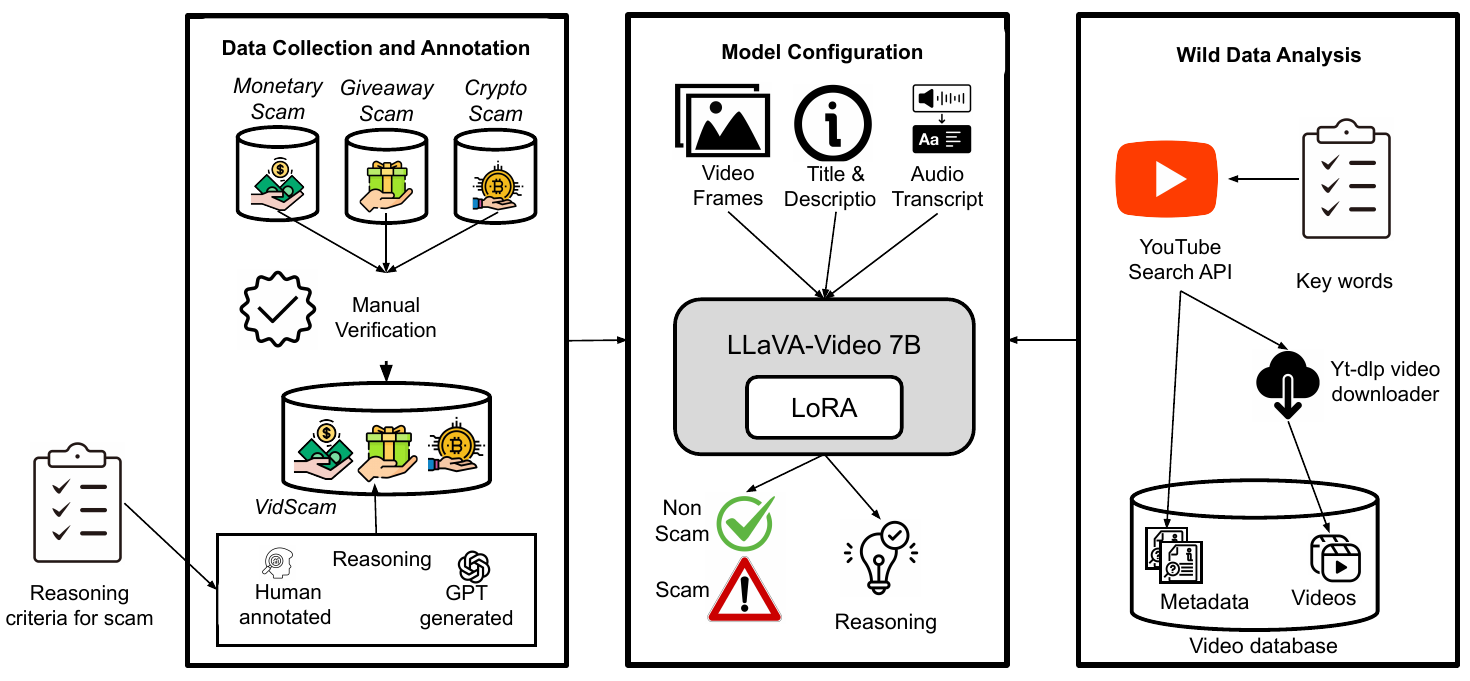}
\caption{Overview: (i) Dataset: Merged and manually re-annotated YouTube scam datasets across three scam types. (ii) Model: Fine-tuned LLaVA-Video-7B fusing frames, titles, and descriptions for scam prediction with reasoning. (iii) Wild Data: \model{} is applied to YouTube videos retrieved via the Search API for at-scale scam detection.} 
\label{fig:method}
\end{figure}

The goal of automated scam video detection on YouTube is to identify content that promotes fraudulent or deceptive activities in accordance with YouTube’s scam policies~\cite{youtube_policy}. To evaluate the incremental value of different information sources, we examine unimodal and multimodal models using textual, audio, and visual content, and propose a multimodal framework that integrates these signals to improve accuracy and provide interpretable, policy-aligned explanations (Figure~\ref{fig:method}). Prior work has explored lightweight baselines based on textual and statistical features. However, textual content has been shown to outperform statistical signals such as views, likes, and comments~\cite{tripathi2022analyzing}. Moreover, these metrics are easily manipulated through fake interactions~\cite{chu2022behind}. Consequently, our study focuses on content-based modalities.

\subsection{Comparative Baselines}
For a comprehensive evaluation of automated scam video detection, we establish comparative baseline models using different combinations of input features and model architectures.

\paragraphb{Modality and Input Configurations.} We explore multiple feature configurations to evaluate how different modalities contribute to scam video detection.

\begin{itemize}[noitemsep,nolistsep,leftmargin=1em]
\item \textbf{Title–Description Based Detector.}
This configuration uses video titles and descriptions, previously shown to be effective for text-based detection~\cite{li2023towards}. Non-English metadata are translated into English via the Google Translation API~\cite{Google-translation-API}, and emojis are converted into textual form using demoji~\cite{demoji}.
\item \textbf{Transcription Based Detector.}
To assess the role of spoken content, we extract audio from each video, generate English transcripts using Whisper~\cite{radford2022robustspeechrecognitionlargescale}, and use them as textual inputs.
\item \textbf{Title–Description–Transcription Based Detector.}
We fuse transcriptions, titles, and descriptions to enrich the textual input.
\item \textbf{Video Based Detector.}
For visual input, we uniformly sample frames from each video to evaluate the contribution of visual cues for scam video detection.
\item \textbf{Video–Transcription Based Detector.}
This configuration combines video frames with corresponding audio transcriptions to jointly capture visual and audio signals.
\item \textbf{Video–Title–Description Based Detector.}
Finally, we combine video frames with processed titles and descriptions to assess the joint effect of textual and visual features on scam detection.
\end{itemize}

\paragraphb{Model Considerations.}
We employ different architectures tailored to each input modality. For textual inputs, we use CryptoScamHunter~\cite{li2023towards}, a state-of-the-art text-based cryptocurrency scam detector implemented as a three-layer feedforward neural network using AllenNLP~\cite{gardner2018allennlpdeepsemanticnatural}. Although designed for crypto scams, its linguistic cues, such as deceptive financial claims, generalize well across scam types. We also train a three-layer feedforward network and a fine-tuned BERT model on all textual configurations, leveraging pretrained language representations for broader generalization across diverse scam types.

For configurations involving visual inputs, we focus on open-source multimodal models to ensure full access to their architectures and training parameters. Closed-source systems like GPT-4V and Gemini, though powerful, pose challenges in cost, scalability, and transparency. We first evaluate two open-source vision–language models, LLaVA-Video-7B and Qwen-2.5-VL-7B, in a zero-shot setting using sampled video frames from the full \dataset{} test set. As shown in Table~\ref{tab:zeroshot}, both models achieve low F1 scores (13–17\%), with Qwen-2.5-VL-7B showing higher precision but poor recall, while LLaVA-Video-7B offers a more balanced trade-off. Additionally, GPT-4o was evaluated on a 100-sample subset of the test set, achieving a higher F1 score of 75.20\%, highlighting the strong zero-shot capability of large proprietary multimodal models. However, these closed models lack fine-tuning flexibility and incur financial costs. Consequently, we select the best-performing open-source model, LLaVA-Video-7B, for fine-tuning to ensure reproducibility and scalability. As we show later, fine-tuned open-source models can surpass large proprietary alternatives.

\subsection{Multimodal Scam Detection with Reasoning}
For our multimodal scam detection framework, \model{}, we consider LLaVA-Video-7B based on its zeroshot performance. Moreover, its diverse video–text pretraining and instruction-tuning further enhance its suitability for reasoning-driven tasks.

\input{tables/zeroshot}
\paragraphb{Multimodal Input Processing.} 
We leverage both textual and visual modalities for YouTube scam video detection. For text, we use video titles and descriptions, translating non-English metadata with the Google Translation API~\cite{Google-translation-API} for consistency. Emojis are converted to text using demoji~\cite{demoji} to preserve semantic meaning. For visuals, we uniformly sample 64 frames from each video. We adopt 64 frames because this is the default pretraining configuration of LLaVA-Video. Increasing to 100 frames yielded only a 0.2\% F1-score gain in the video-only setting, while tripling runtime and increasing memory usage by 1.35 times. In the multimodal setting, 100 frames caused the combined inputs (frames + title + description) to exceed context limit, making it infeasible for many videos. In LLaVA-Video, text is tokenized with the Qwen-2 tokenizer~\cite{yang2024qwen2technicalreport}, and frames are encoded using the SigLIP vision encoder~\cite{zhai2023sigmoid}. The fused text–vision representations are processed by the language model to produce the predicted scam label and reasoning.

\paragraphb{Supervised Finetuning with LoRA.}
We formulate YouTube scam detection as a multimodal instruction-following task~\cite{shengyu2023instruction}. Each training instance includes the video’s visual frames, temporal metadata (duration and frame timestamps), and its title and description, paired with an instruction prompt: \textit{“Is this a scam video? Answer Yes/No and explain your reasoning. Analyze the content and refer to the official YouTube Terms and Conditions violations for scams (List of scam criteria).”} The model is trained to output both a binary prediction (Yes/No) and a policy-aligned textual rationale (Section~\ref{sec:dataset}).

To efficiently adapt LLaVA-Video-7B, we apply Low-Rank Adaptation (LoRA)~\cite{hu2022lora}, a parameter-efficient fine-tuning technique that inserts small trainable matrices into select layers while freezing the rest. This approach reduces computational and memory costs, enabling domain-specific adaptation without compromising performance. In our setup, we use a LoRA rank of 128 and an alpha of 32, adding only 55.71 MB of trainable parameters, which is about 0.69\% of the model’s total size (8086.05 MB). The model is trained for up to 10 epochs with a learning rate of $1e^{-3}$, resulting in an efficient yet effective adaptation for scam detection.

\paragraphb{Reasoning Data Augmentation for Training.} \label{sec:gpt_generation}
To fine-tune the model for generating policy-grounded reasoning, we used 200 human-annotated samples from Section~\ref{sec:human_annotation}, splitting them evenly between training and evaluation. Fine-tuning on this small set resulted in limited performance (55.71\% F1), underscoring the need for additional training data. To scale up, we employed GPT-4o Vision to generate reasoning aligned with YouTube’s policy criteria under two input settings: (i) title, description, and frames, and (ii) frames only. The latter produced reasoning more consistent with human annotations. The GPT-generated explanations achieved a BERTScore~\cite{zhang2019bertscore} of 0.88 compared to 100 human-annotated validation samples (60 scam, 40 non-scam), indicating high semantic similarity. Moreover, we manually evaluated quality of the generated reasoning with the validation set. For scam videos, GPT-generated reasoning fully aligned in 53 cases and partially aligned in 7. For non-scam videos, GPT was instructed to produce brief summaries indicating legitimacy and it generated reliable summaries for all 40 cases. We then augmented \dataset{} with GPT-generated reasoning for 1,300 additional videos (prompt used is available in Appendix~\ref{Appendix:2}). An ablation study on training data size (see Figure~\ref{fig:ablation_training_data}) further shows that performance steadily improves as more reasoning data are added. We use this final model configuration for all subsequent evaluations in the paper.

\input{figures/ablation_data_size}

\subsection{Implementation Details}
All models were implemented in Python. Text-only experiments (the three-layer neural network and pretrained BERT) were conducted using the AllenNLP framework~\cite{gardner2018allennlpdeepsemanticnatural} on a system with an NVIDIA RTX 4090 GPU. Experiments involving the LLaVA-Video model were run on a machine with an NVIDIA A100 GPU.

%% file: tables/zeroshot.tex
\begin{table}[!t]
\resizebox{\linewidth}{!}{
\begin{tabular}{lcccc}
\hline
Model          & \multicolumn{1}{l}{Acc.\%} & \multicolumn{1}{l}{Prec.\%} & \multicolumn{1}{l}{Rec.\%} & \multicolumn{1}{l}{F1 Score\%} \\ \hline
LLaVA-Video-7B & 76.78                        & 27.4                          & 12.69                      & 17.35                        \\
Qwen-2.5-VL-7B & 80.41                        & 40.22                         & 8.45                       & 13.96                        \\ 
GPT-4o Vision$^{\star}$ & 69.00  & 73.44 & 77.05 & 75.20 \\
\hline
\end{tabular}
}
\caption{Zero shot performance of vision-language models. ($^{\star}$Evaluated on a random subset consisting 100 videos)}
\label{tab:zeroshot}
\end{table}

%% file: figures/ablation_data_size.tex
\begin{figure}[!t]
\centering
\resizebox{0.85\linewidth}{!}{
\begin{tikzpicture}
\begin{axis}[
    width=\linewidth, height=4.5cm,
    xlabel={Training Samples},
    ylabel={Percentage},
    ymin=40, ymax=100,
    grid=major,
    xtick={100,200,500,750,1000,1400},
    tick label style={font=\scriptsize},
    label style={font=\scriptsize},
    legend style={
        font=\footnotesize,
        at={(0.98,0.02)}, anchor=south east
    }
]


\addplot+[mark=triangle, thick, blue] coordinates {
    (100,43.41) (200,47.44) (500,55.18) (750,63.88) (1000,68.26) (1400,76.16)
};
\addlegendentry{Precision}

\addplot+[mark=square, thick, red] coordinates {
    (100,77.73) (200,88.64) (500,91.31) (750,86.64) (1000,92.43) (1400,91.09)
};
\addlegendentry{Recall}

\addplot+[mark=diamond, thick, green!60!black] coordinates {
    (100,55.71) (200,61.8) (500,68.79) (750,73.53) (1000,78.52) (1400,82.96)
};
\addlegendentry{F1 Score}

\end{axis}
\end{tikzpicture}}
\caption{Effect of training size on performance.}
\label{fig:ablation_training_data}
\end{figure}

%% file: sections/result.tex
\input{tables/main_res}

Our experiment results depict performance differences across the text-only, visual-only, and multimodal detection models of YouTube scam videos. Table \ref{tab:main_res} summarizes the classification performance across different input modalities and feature configurations on the \dataset{} test set. These results form the foundation for the following subsections, where we analyze the performance in detail.

\input{tables/sota_gap}

\subsection{Performance Under Different Modality}

\paragraphb{Scam Detection using Text-Only Modality.}
We first evaluate textual features, video titles and descriptions, for scam detection. The state-of-the-art text-based model CryptoScamHunter~\cite{li2023towards} achieves 92.03\% F1 score on its cryptocurrency-focused dataset but drops to 46.88\% F1 score on \dataset{}, which covers broader scam types (as shown in Table~\ref{tab:sota_gap}). This gap highlights the model’s limited generalizability, expected given its simple three-layer feedforward design and narrow training scope.
Fine-tuning the same network on \dataset{} improves F1 score to 73.46\%, though precision declines due to increased false positives. Fine-tuning BERT on the same dataset further boosts performance to 76.61\% F1 score.

We next assess audio transcriptions for scam detection. Models using transcriptions alone perform worse, achieving 66.39\% F1 with a three-layer network and 71.76\% with fine-tuned BERT, reflecting that many scam videos contain only background music or non-informative audio. In our annotated subset of 200 videos, 90 contain only background music, including 74 scam videos, for which Whisper produces generic or meaningless transcripts with no scam-relevant cues.
Finally, combining all three textual inputs—titles, descriptions, and transcriptions—yields the strongest results. The three-layer network reaches 77.30\% F1 score (vs. 73.46\% with titles+descriptions, 66.39\% with transcriptions), while fine-tuned BERT achieves 77.98\%, showing modest gains from integrating transcriptions.

\paragraphb{Scam Detection using Visual Modality.}
 We evaluate the effectiveness of visual information using frame-level features extracted from video content. Finetuning the LLaVA-Video-7B model on video frames alone achieves an F1 score of 79.61\%, demonstrating that visual cues can reliably identify scam content. The visual-only model also attains higher precision (76.81\%) than text-only models, indicating fewer false positives and highlighting the robustness of visual features for scam detection. 

\paragraphb{Scam Detection using Multiple Modalities.}
We evaluate multimodal performance by combining video frames with textual inputs (titles, descriptions, and transcriptions). Fine-tuning the multimodal models produces consistent gains across all configurations (as highlighted in Table~\ref{tab:main_res}). Combining frames with transcriptions achieves an F1 score of 79.49\%, comparable to the frame-only model (79.61\%). However, integrating frames with titles and descriptions yields the best overall result (80.62\% F1 score), surpassing both the best text-only model (77.98\% F1 score) and the visual-only model, with a 5.10\% recall improvement. Overall, adding visual cues to textual metadata improves the F1 score by 4.01\%. These findings indicate that textual features from titles and descriptions complement visual signals, enhancing YouTube scam video detection. Moreover, unimodal models are more vulnerable to adversarial perturbations (Appendix~\ref{Appendix:text_limitattion}), reinforcing the robustness of multimodal learning.

\input{tables/ml_comparision}

\subsection{Policy Grounded Multimodal Detector}
The proposed multimodal scam detection framework, \model{}, achieves the best overall performance, reaching an F1 score of 82.96\% after instruction tuning with video frames, titles, descriptions, and policy-grounded reasoning data. Incorporating policy-grounded reasoning yields a 2.34\% improvement over the non-reasoning multimodal baseline (80.62\% F1 score), indicating that explicit reasoning helps the model better capture nuanced scam indicators. This gain comes with minimal overhead, increasing tokens by only 0.85\%. In addition, the generated explanations closely match human-annotated reasoning, achieving a BERTScore of 0.89 on 100 benchmark samples (Section~\ref{sec:gpt_generation}). Beyond accuracy gains, policy-aligned reasoning enhances transparency and consistency with YouTube’s scam criteria.


\begin{tcolorbox}[colback=lightgray,colframe=lightgray, boxsep=1pt,left=2pt,right=2pt,top=0pt,bottom=0pt]
\textbf{Takeaway:} Multimodal models outperform unimodal ones in YouTube scam detection, with text providing strong cues, visuals improving precision, and policy-grounded reasoning enhancing both performance and interpretability.
\end{tcolorbox}

\subsection{Comparison with Existing ML Approaches}
Prior work~\cite{tripathi2022analyzing} explored lightweight models, including SVC, AdaBoost, and Random Forest, for detecting scam videos using titles and descriptions. As shown in Table~\ref{tab:ml_gap}, these methods achieve moderate performance (F1 score up to 72\%) but have limited robustness to adversarial manipulation.

To evaluate realistic evasion strategies, we test textual attacks including leet-style obfuscations (e.g., replacing characters with visually similar symbols such as “Fr33 G!ftC@rd”) and semantic perturbations generated using TextAttack~\cite{morris2020textattack}, modeling both character-level and meaning-preserving manipulations commonly observed in scam content on a sample of 100 scam videos from the test set. Under semantic perturbations, prior lightweight baselines show accuracy drops of 12--38\% (Table~\ref{tab:ml_gap}). Similarly, the text-only BERT model is highly susceptible to manipulation, suffering a 33\% accuracy decrease under leet transformations and a 15\% drop under semantic edits. A representative example (Appendix~\ref{Appendix:text_limitattion}) illustrates how text obfuscation causes text-only models to misclassify scams that remain visually identifiable through cues such as fake code generators and verification prompts.

We further examine robustness to visual perturbations by applying Gaussian noise, rotation, and blur to video frames using standard image transformation tools~\cite{natlee2024blurgenerator}. The visual-only LLaVA model shows relative robustness to Gaussian noise and rotation, with accuracy declines of only 3\% and 5\%, respectively, but degrades substantially under blur (30\%), indicating reliance on fine-grained visual details. In contrast, the proposed multimodal model exhibits substantially improved resilience, with textual perturbations reducing accuracy by only 1\% and visual transformations degrading performance by at most 3\%. Overall, these findings demonstrate that while lightweight text-based baselines offer computational efficiency, multimodal fusion is essential for robust scam detection under adversarial conditions.

\subsection{Computational Cost Analysis}
To assess computational efficiency, we measure inference time and GPU memory usage across model configurations (details available in Appendix~\ref{Appendix:model_cost}). The three-layer neural network is most efficient, using only 0.10 GB of memory and 1 ms per sample with all textual inputs. In comparison, BERT requires 1 GB and 0.13 s per sample, reflecting the added cost of transformer-based models.
Multimodal models introduce significantly higher overhead. LLaVA-Video, when processing only video frames, consumes 55.6 GB of memory and 2 s per sample. Adding modalities further increases usage to 66.8 GB and 2.1 s for frames with transcriptions, and 57.1 GB and 2.0 s for frames with titles and descriptions. Overall, these results underscore the trade-off between accuracy and computational cost in multimodal detection systems. However, for platforms like YouTube, such computational resources are quite feasible.

\input{figures/qualitative}

%% file: tables/main_res.tex
\begin{table*}[t!]
\centering


\resizebox{0.9\linewidth}{!}{
\begin{tabular}{llcccc}
\toprule
Training Data                                      & Model          & \multicolumn{1}{l}{Accuracy\%} & \multicolumn{1}{l}{Precision\%} & \multicolumn{1}{l}{Recall\%} & \multicolumn{1}{l}{F1 Score\%} \\ \midrule
\multirow{2}{*}{Title, Description}                & 3Layer NN      & 88.20                         & 64.64                         & 85.08                      & 73.46                        \\
                                                   & Bert           & 90.21                        & 70.75                         & 83.52                      & 76.61                        \\ \hline
\multirow{2}{*}{Transcription}                     & 3Layer NN      & 86.15                        & 62.14                         & 71.21                      & 66.39                        \\
                                                   & Bert           & 87.99                        & 65.38                         & 79.51                      & 71.76                        \\ \hline
\multirow{2}{*}{Title, Description, Transcription} & 3Layer NN      & 90.94                        & 74.43                         & 80.40                       & 77.30                         \\
                                                   & Bert           & 90.68                        & 71.35                         & 85.97                      & 77.98                        \\ \hline
Video Frames                                       & Llava-Video-7B & 91.88                        & 76.81                         & 82.63                      & 79.61                        \\ \hline
Video Frames, Transcription                       & Llava-Video-7B & 91.11                        & 71.33                         & 89.76                      & 79.49                        \\
Video Frames, Title, Description                  & Llava-Video-7B & 91.45                        & 71.36                         & 92.65                      & 80.62                        \\ 

\hline

Video Frames, Title, Description, Reasoning       & Llava-Video-7B & \textbf{92.82}               & \textbf{76.16}                & \textbf{91.09}             & \textbf{82.96}               \\ \bottomrule
\end{tabular}
}
\caption{Comparison of the proposed \model{} with existing scam detection models across different modalities.}
\label{tab:main_res}
\end{table*}

%% file: tables/sota_gap.tex
\begin{table}[!t]
\centering

\Huge
\resizebox{\linewidth}{!}{
\begin{tabular}{lcccc}
\toprule
Dataset & Acuracy \% & Precision \% & Recall \% & F1 score \% \\ \midrule
\crypto{}  & 94.36   & 95.98     & 88.40  & 92.03    \\
\dataset{}  & 86.53   & 96.53     & 30.96  & 46.88    \\ \bottomrule
\end{tabular}
}
\caption{Performance of state-of-the-art text model}
\label{tab:sota_gap}
\end{table}

%% file: tables/ml_comparision.tex
\begin{table}[!t]
\centering
\Huge
\resizebox{\linewidth}{!}{
\begin{tabular}{lccccc}
\toprule
Model         & Acc. \% & Prec. \% & Rec. \% & F1 score \% & \begin{tabular}[c]{@{}l@{}}Acc. drop \%\end{tabular} \\ \midrule
SVC           & 84.35       & 56.02        & 85.96     & 67.83       & 12.00                                                                                 \\
Adaboost      & 87.98       & 66.34        & 75.94     & 70.82       & 22.00                                                                                 \\
RF & 90.59       & 81.36        & 66.14     & 72.97       & 38.00                                                                                 \\ 
\model{} & 92.82    & 76.16  & 91.09    & 82.96    & 1.00 \\
\bottomrule
\end{tabular}
}
\caption{Comparison with Existing ML approaches}
\label{tab:ml_gap}
\end{table}

%% file: figures/qualitative.tex
\begin{figure*}[!th]
    \centering
    \includegraphics[width=0.9\linewidth]{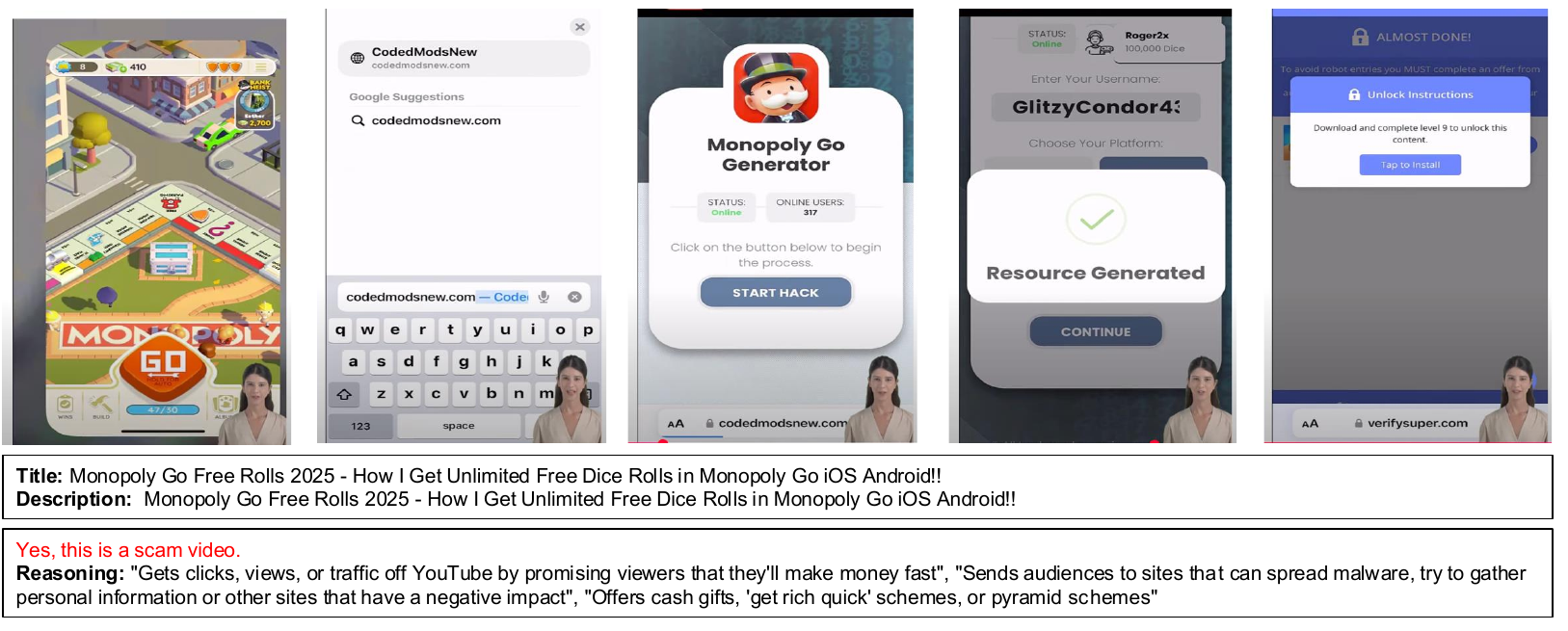}
    \vspace{0.8em} 
    \includegraphics[width=0.92\linewidth]{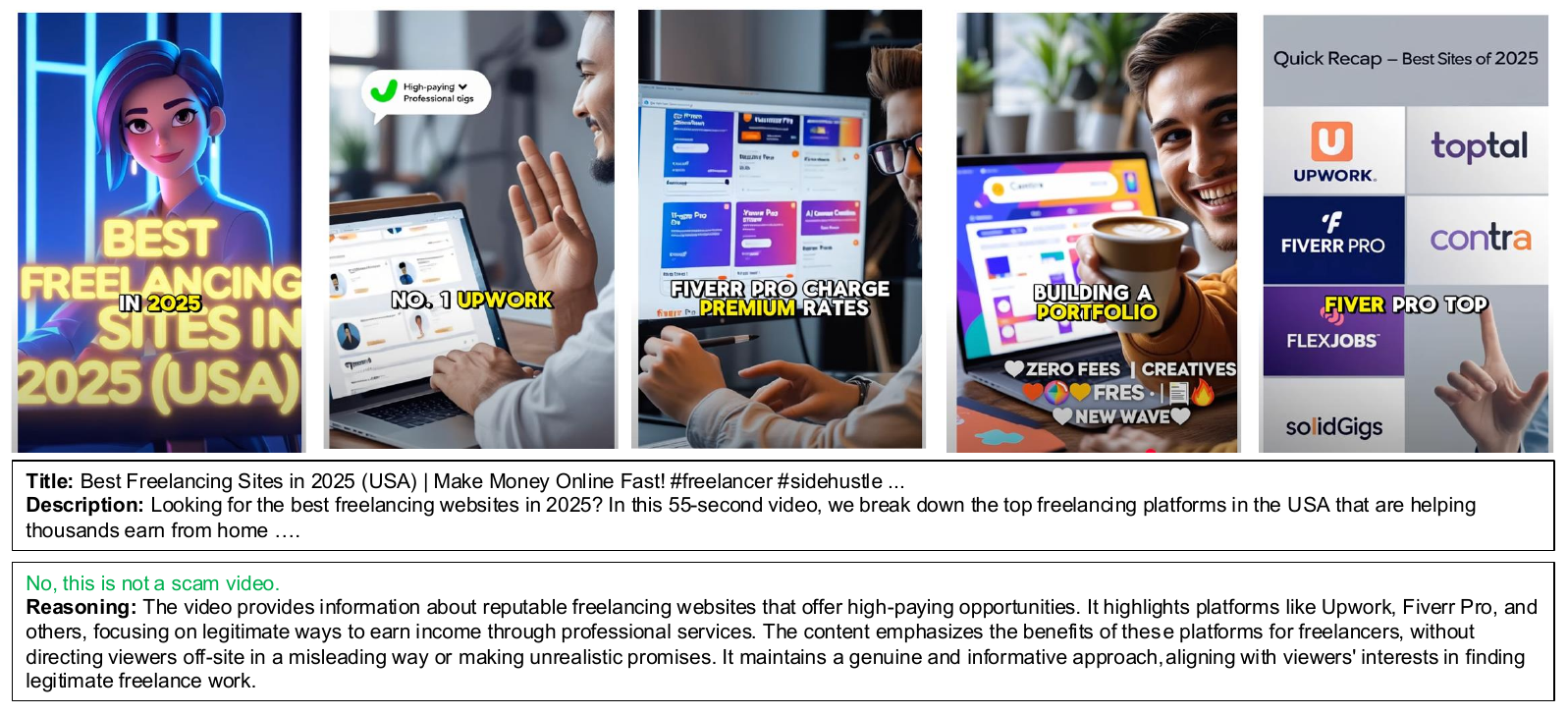}
    \caption{Examples classified by the multimodal framework with policy-grounded reasoning. (first: scam; second: non-scam)}
    \label{fig:qualitative}
\end{figure*}

%% file: sections/wild_data.tex
We further evaluate our YouTube scam video detection framework on real-world (“in-the-wild”) YouTube videos. To construct this dataset, we curated a fixed set of keywords to increase the likelihood of retrieving scam-related content. These keywords were drawn from prior studies: Bouma-Sims et al.~\cite{bouma2021first} identified query patterns targeting giveaway scams, Tripathi et al.~\cite{tripathi2022analyzing} used Google Trends to extract monetary scam–related terms, and Li et al.~\cite{li2023towards} compiled keywords linked to cryptocurrency scams. In total, we collected 70 keywords (listed in Appendix~\ref{Appendix:keyword}: Table~\ref{tab:keywords}). Using these, we queried the YouTube Data API~\cite{YouTube_search} to retrieve video IDs and metadata (titles and descriptions) and used the \texttt{yt-dlp}~\cite{yt-dlpgit} library to download the videos. Due to API and availability constraints, some videos could not be retrieved. Running the pipeline daily over three months produced a final dataset of 6,374 videos.

We then deploy the best-performing multimodal model, \model{} (integrating video frames, titles, and descriptions), to analyze the wild dataset. Out of 6,374 YouTube videos, \model{} flagged 1,864 as scams and produced policy-grounded rationales supporting each classification. Noteably, \emph{\textbf{316 of the videos flagged as scams by \model{} have since been removed from YouTube}}, underscoring both the practical relevance of our detection framework and its potential alignment with platform moderation outcomes.

\input{tables/wild_criteria_dist}

Table~\ref{tab:wild_scam_distribution} shows the distribution of scam criteria in the detected videos. The most common scams involved cash gifts and get-rich-quick schemes (1,711), followed by redirections to malicious sites (1,447) and fast-money promises (1,361). Less frequent but still notable were unbounded giveaways (257), off-site redirections via misleading content (237), and impersonation of individuals or organizations (65). Extremely rare cases included videos claiming to commit crimes on behalf of users (7). These findings demonstrate that multimodal detection effectively uncovers a wide range of scam tactics, with financial deception and malicious redirection dominating real-world occurrences. 

\input{tables/wild_data_eval}
To further evaluate detection performance and reasoning quality, we manually inspected 100 randomly selected videos (50 scams and 50 non-scams). The model misclassified 7 videos, achieving an overall accuracy of 93\%, with 94\% precision and 92.16\% recall. The reasoning outputs for correctly classified videos were rated on a three-level scale. ‘Fully Aligned’: accurately captures the key scam criteria in line with human judgment; ‘Partially Aligned’: identifies some relevant criteria but misses some or includes unrelated criteria; ‘Incorrect’: fails to generate the any relevant scam criteria. As shown in Table~\ref{tab:wild_data_evaluation}, 72 had fully aligned reasoning, 16 were partially aligned, and 1 had no generated reasoning (labeled Incorrect). These results indicate that when the model predicts correctly, its reasoning is generally reliable and closely matches human annotations. Representative videos with generated reasoning are provided in Figure~\ref{fig:qualitative} and in Appendix~\ref{Appendix:qualitative}.



\begin{tcolorbox}[colback=lightgray,colframe=lightgray, boxsep=1pt,left=2pt,right=2pt,top=0pt,bottom=0pt]
\textbf{Takeaway:} Our multimodal framework detected 1,864 real-world scam videos with policy-aligned reasoning; financial deception and malicious redirection were most common, and 316 videos were removed subsequently.
\end{tcolorbox}

%% file: tables/wild_criteria_dist.tex
\begin{table}[!t]
\centering
\small
\resizebox{\linewidth}{!}{
\begin{tabular}{cccccccc}
\toprule
Crit. 1 & Crit. 2 & Crit. 3 & Crit. 4 & Crit. 5 & Crit. 6 & Crit. 7 \\
\midrule
7 & 257 & 237 & 1361 & 1447 & 1711 & 65 \\

\bottomrule
\end{tabular}
}
\caption{Distribution of scam criteria in the wild dataset}
\label{tab:wild_scam_distribution}
\end{table}

%% file: tables/wild_data_eval.tex



\begin{table}[!t]
\centering
\resizebox{\linewidth}{!}{
\begin{tabular}{lcccc}
\toprule
\multirow{2}{*}{Label} & \multicolumn{3}{c}{\makecell{Correctly classified \\(Reasoning Quality)}} & \multirow{2}{*}{\makecell{Misclassified}} \\ \cline{2-4}
 & \makecell{Fully \\ aligned} & \makecell{Partially \\ aligned} & \makecell{Incorrect} & \\ \midrule
Scam     & 33 & 13 & 1 & 3 \\
Non scam & 43 & 3  & 0 & 4 \\ \hline
Total    & 76 & 16 & 1 & 7 \\ \bottomrule
\end{tabular}
}
\caption{Manual evaluation of generated reasoning (n=100)}
\label{tab:wild_data_evaluation}
\end{table}

%% file: sections/discussion.tex
Our study shows that multimodal modeling, combining video, text, and policy-grounded reasoning, outperforms traditional metadata-based approaches for detecting YouTube scams. In addition to higher accuracy, it provides policy-aligned explanations, enhancing transparency and explainability in automated moderation.

\paragraphb{Integration within Existing Detection Pipelines.} The average inference time of our model is 2 seconds and memory usage is  57.1 GB on an A100 GPU. Several standard optimization techniques can further reduce deployment overhead. For example, frame preprocessing lowers average inference time by approximately 1 second, while FP8 quantization reduces memory usage by about 50\%. Additional methods such as knowledge distillation, batch processing, and more aggressive quantization can be applied for deployment to further optimize the cost. Moreover, the text based detectors can still serve as inexpensive, high-recall filters in a cascaded deployment setting. In practice, lightweight text models can be used solely to route videos rather than make final decisions, allowing the system to eliminate a large fraction of benign content before invoking more expensive multimodal models.

\paragraphb{Limitations and Future Work.}
Despite these contributions, our work has a few limitations. First, although our dataset is large and diverse, it cannot fully capture the evolving landscape of scams, relying on videos available from prior literature. Second, we analyze uniformly sampled frames rather than full video sequences, potentially missing temporal cues that distinguish scams from legitimate content. Third, some reasoning data is GPT-augmented, which may not fully capture nuanced policy interpretations. Fourth, while open-source models like LLaVA-Video enable cost-effective fine-tuning, commercial models may achieve higher performance, so our results primarily establish an improved baseline. Finally, our focus is limited to YouTube, and findings may not generalize to other platforms, though the framework is adaptable.

Future work could extend this framework to cross-platform settings by adapting the criteria to platform specific policies and collecting datasets. Additionally, incorporating temporal modeling of full video streams to capture time-dependent scam tactics, develop automated frame selection or temporal grounding for improved efficiency and accuracy, and conduct cross-platform studies to uncover broader scam strategies and inform coordinated defenses.


\paragraphb{Ethical Considerations.} Our study uses publicly available YouTube videos and is exempt from IRB review. We also adhered to YouTube API rate limits during data collection to minimize network impact, and reported identified scam videos to YouTube for further review. 

%% file: sections/conclusion.tex
In this work, we conducted a systematic study of scam detection on YouTube through a multimodal lens. Unlike prior approaches that primarily rely on textual or statistical metadata, our approach leverages a combination of video frames, title, description, and policy-grounded reasoning to effectively identify scam videos on YouTube. We constructed a multimodal dataset of YouTube scam content with associated reasoning criteria and demonstrated the effectiveness of our approach. Applying our best-performing model, \model{}, to a large-scale wild dataset, we uncovered thousands of scam videos, highlighting both the prevalence of scams on the platform and the practical potential of multimodal, policy-aligned detection frameworks.

%% file: sections/appendix.tex
\onecolumn

\section*{Appendix}

\section{Qualitative Results}\label{Appendix:qualitative}
\input{figures/qualitative_apendix}
\clearpage
\section{Krippendorf's alpha scores}\label{Appendix:1}
\input{tables/k_alpha}

\section{Keywords list for Wild Data Collection}\label{Appendix:keyword}
\input{tables/keywords}

\section{Computational Cost} \label{Appendix:model_cost}
\input{tables/cost}

\clearpage
\section{Quality of the Datasets} \label{Appendix:data_quality}
\input{tables/data_validation}
\section{Textual obfuscation Example}\label{Appendix:text_limitattion}
\input{sections/adversarial}
\section{GPT Assisted Reasoning Generation Prompt}\label{Appendix:2}
\input{tables/prompt_scam}
\clearpage
\input{tables/prompt_nonscam}











%% file: figures/qualitative_apendix.tex
\begin{figure*}[!th]
    \centering
    \includegraphics[width=\linewidth]{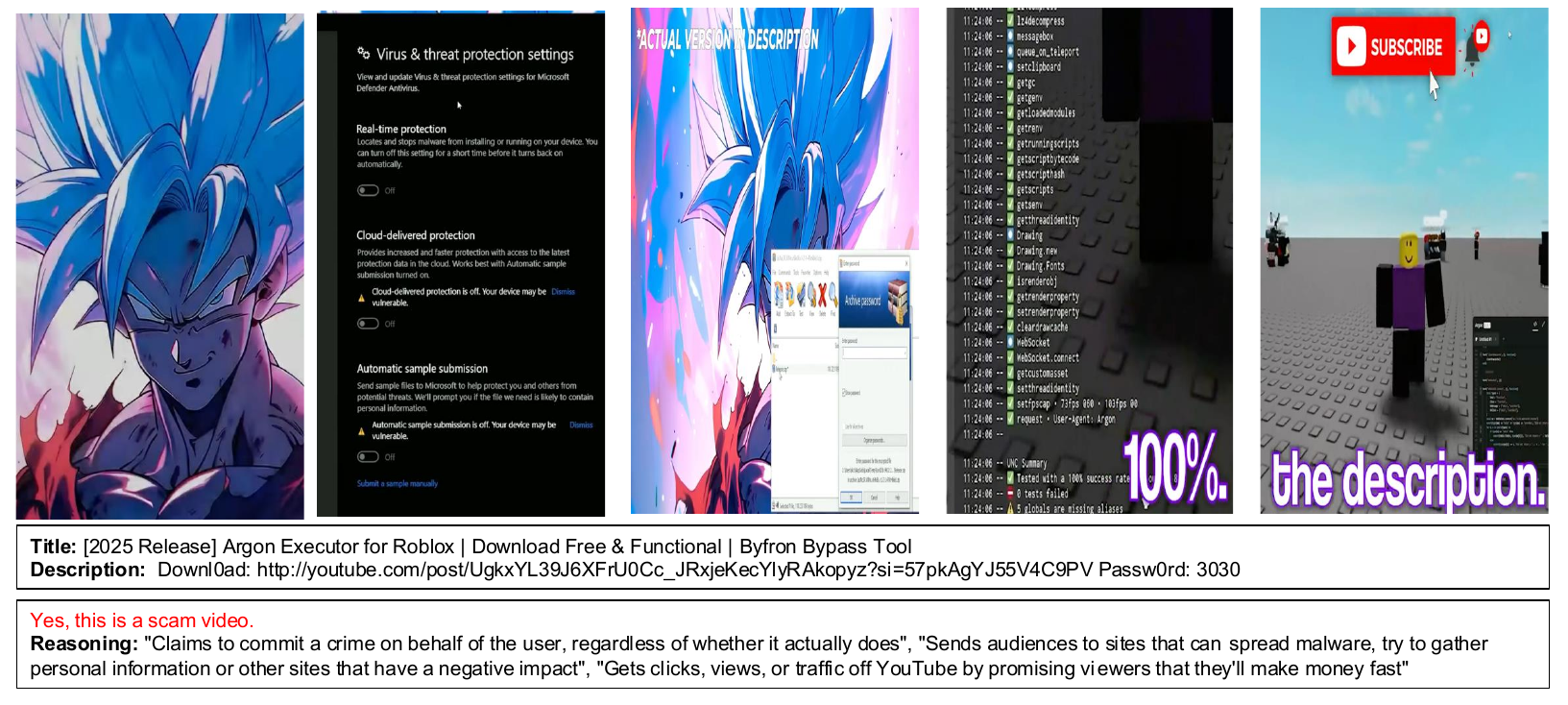}
    \vspace{0.8em} 
    \includegraphics[width=\linewidth]{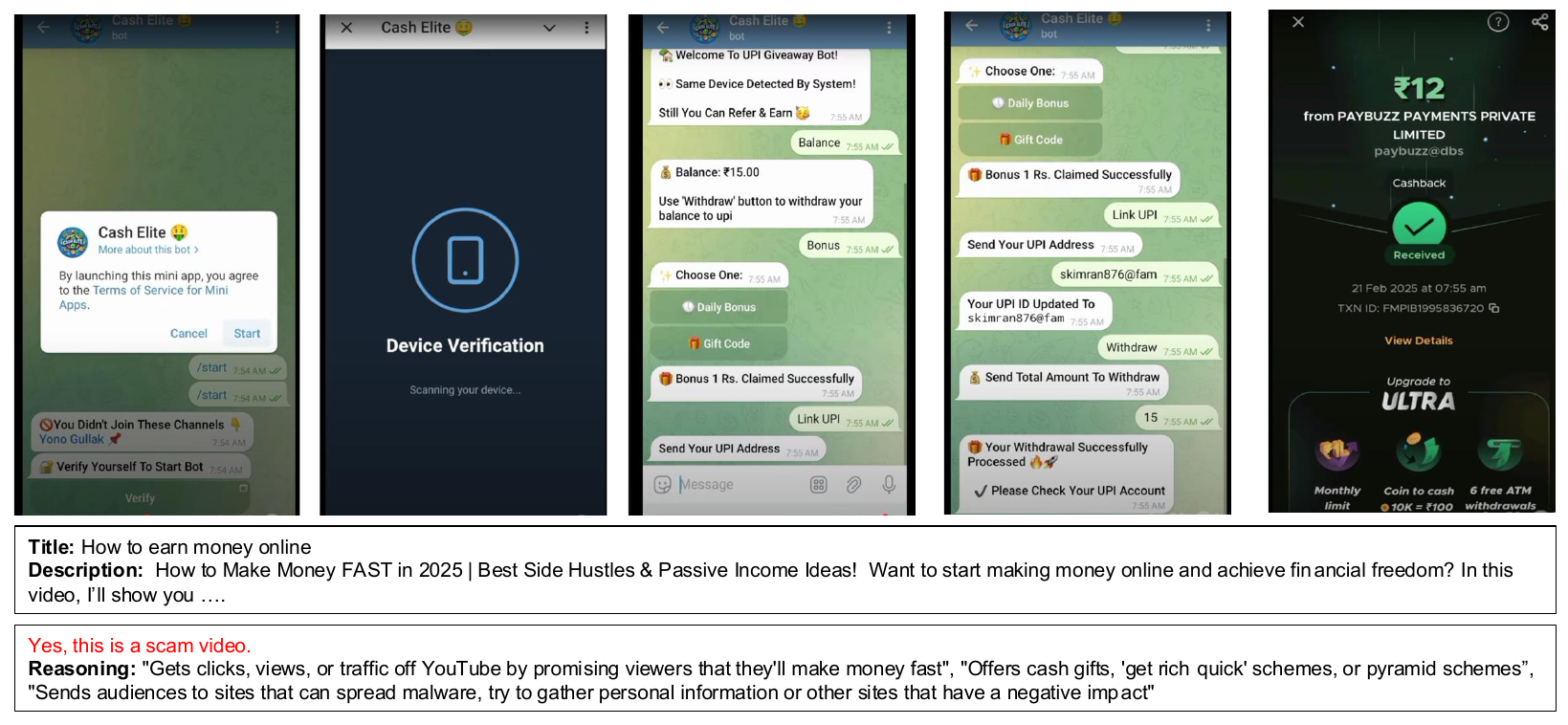}
    \caption{Examples classified by the proposed multimodal framework with policy-grounded reasoning.}
    \label{fig:qualitative_appendix}
\end{figure*}

%% file: tables/k_alpha.tex
\begin{table}[h]
\centering

\resizebox{\linewidth}{!}{
\begin{tabular}{lcccccc}
\toprule
                                   &                                    & \multicolumn{5}{c}{Column Name}                                                                                                                                                                                                      \\ \cline{3-7} 
\multirow{-2}{*}{\makecell{Training \\Session}} & \multirow{-2}{*}{\makecell{Number of \\ Sample}} & Agree with Ground Truth? & Label (Scam, Non scam) & Broad Scam Crit. & Narrow Scam Crit. & Modality \\ \midrule
1                                  & 10                                 & 0.54                                             & 0.70                                                           & 0.40                                   & 0.68                                 & 0.02                             \\
2                                  & 10                                 & 1.00                                             & 1.00                                                           & 1.00                                   & -                                    & -                                \\
3                                  & 19                                 & -0.02                                            & 0.73                                                           & 0.39                                   & -0.11                                & 0.08                             \\
4                                  & 10                                 & 0.00                                             & 0.85                                                           & 0.29                                   & 0.52                                 & 0.31                             \\
5                                  & 10                                 & 0.29                                             & 0.87                                                           & 0.44                                   & 0.44                                 & 0.31                             \\
6                                  & 16                                 & 0.78                                             & 0.90                                                           & 0.67                                   & 0.52                                 & 0.33                             \\
7                                  & 10                                 & 0.62                                             & 0.80                                                           & 0.77                                   & 0.89                                 & 0.89                             \\
8                                  & 15                                 & 0.78                                             & 0.91                                                           & 0.83                                   & 1.00                                 & 0.57                             \\
9                                  & 15                                 & 0.73                                             & 0.91                                                           & 0.83                                   & 0.72                                 & 0.70                             \\ \bottomrule
\end{tabular}
}
\caption{Krippendorf's alpha scores after each iteration}
\label{tab:k_alpha}
\end{table}

%% file: tables/keywords.tex
\begin{table*}[h]
\centering
\resizebox{\linewidth}{!}{
\begin{tabular}{p{0.18\linewidth} p{0.78\linewidth}}
\hline
\textbf{Category} & \textbf{Keywords}                                                                                                                                                                                                                                                                                                                                                                                                                                                                                                                                                                                                                                                                                                                                                                                                                                                                                                                                                                                                                                                                    \\ \hline
Giveaway Scam                          & Free Amazon.com eGift Card, Free Visa Gift Card, Free DoorDash eGift Card,Free Sephora Gift Card, Free Razer Gold eGift Card, Free Uber eGift Card, Free Starbucks eGift Card, Free Mastercard Gift Card, Free Starbucks Gift Card, Free Ulta Beauty Gift Card, Free Netflix eGift Card, Free Visa Virtual eGift Card, Free Spotify Premium eGift Card, Free Apple Gift Card, free honor of kings currency, free Last war: survival currency, free whiteout survival currency, free royal match currency, free pubg mobile currency, free monopoly go currency, free candy crush saga currency, free pokemon tcg pocket currency, free coin master currency, free roblox currency, apple tech support, microsoft tech support, nvidia tech support, samsung tech support, sony tech support, intel tech support, dell tech support, panasonic tech support, UBS support, Morgan Stanley support, Bank of America support, J.P. Morgan, Private Bank support, Citigroup support, BNP Paribas support, Goldman Sachs support, Julius Baer support, Raymond James support, HSBC support \\ \hline
Monetary Scam                          & How to earn money online, ways to earn money online, earn money online fast, best way to earn money online, how to make money as kid, how to make money online with ai, how to make money online as teen                                                                                                                                                                                                                                                                                                                                                                                                                                                                                                                                                                                                                                                                                                                                                                                                                                                                             \\ \hline
Cryptocurrency Scam                    & passive income, huge profit, easy profit, building wealth with crypto, earn free bnb, earn free eth, UniSwap, SushiSwap, PancakeSwap, Aave, Avalanche/Avax, Polygon/Matic, Fantom/FTM, arbitrage bot, front-running bot, flashloan bot, MEV bot, snipe Bot, trading bot, DeFi bot                                                                                                                                                                                                                                                                                                                                                                                                                                                                                                                                                                                                                                                                                                                                                                                                    \\ \hline
\end{tabular}
}
\caption{Keywords used for wild data collection}
\label{tab:keywords}
\end{table*}

%% file: tables/cost.tex
\begin{table}[h]
\centering

\label{tab:cost}
\begin{tabular}{llcc}
\hline
Model       & Input Feature                     & Inference Time (sec/sample) & Memory (GB) \\ \hline
3 layer NN  & Title, Description                & 0.0004                                          & 0.08                            \\
Bert        & Title, Description                & 0.09                                            & 0.99                            \\
3 layer NN  & Transcription                     & 0.001                                           & 0.09                            \\
Bert        & Transcription                     & 0.10                                             & 1.00                               \\
3 layer NN  & Title, Description, Transcription & 0.001                                           & 0.10                             \\
Bert        & Title, Description, Transcription & 0.13                                            & 1.00                               \\
LLaVA-Video & Video Frames                      & 2.00                           & 55.58           \\
LLaVA-Video & Video Frames, Transcription       & 2.09                            & 66.81          \\
LLaVA-Video & Video Frames, Title, Description  & 2.05                            & 57.12          \\ \hline
\end{tabular}
\caption{Memory usage and average inference time for each model.}
\end{table}

%% file: tables/data_validation.tex
\begin{table}[h]
\centering

\begin{tabular}{lccc}
\toprule
Dataset                      & \begin{tabular}[c]{@{}c@{}}\# Scam in \\ Ground Truth\end{tabular} & \begin{tabular}[c]{@{}c@{}}\# Relabeled \\ Non-scam\end{tabular} & \% Mislabeled \\ \midrule
\monetary{} & 278                                                                & 9                                                                  & 3.24          \\
\giftcard{} & 146                                                                & 5                                                                  & 3.42          \\
\crypto{}   & 580                                                                & 41                                                                 & 7.07          \\ \bottomrule
\end{tabular}
\caption{Quality of ground truth labels in each dataset.}
\label{tab:data_validation}
\end{table}

%% file: sections/adversarial.tex
\begin{figure*}[h]
\centering
\includegraphics[width=\textwidth]{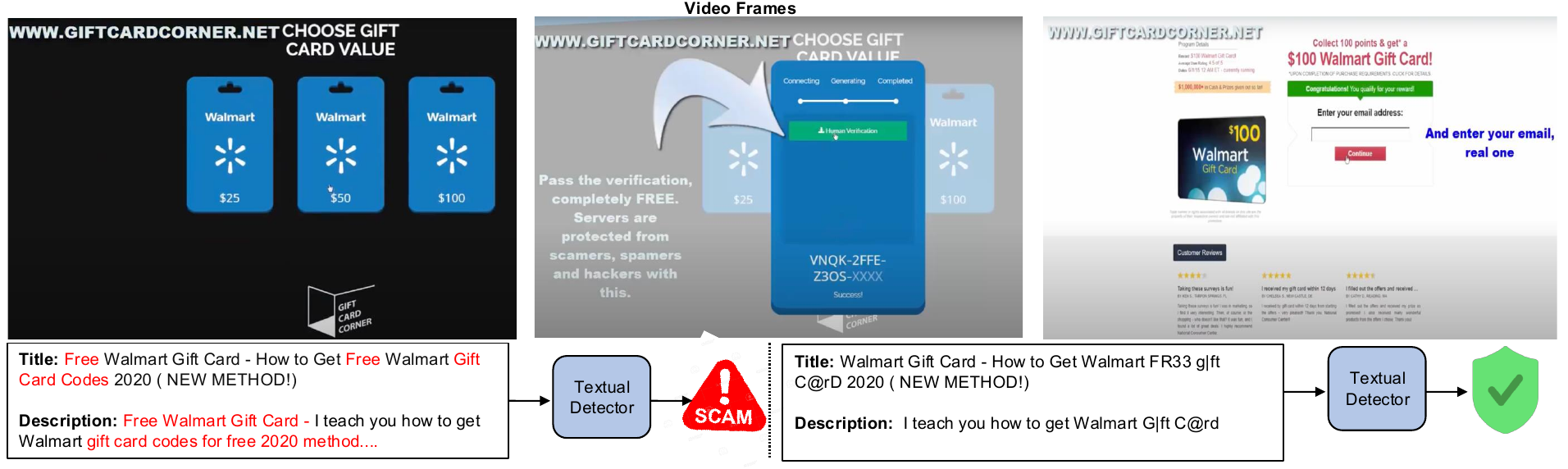}
\caption{Effect of textual obfuscation: The fine-tuned BERT model accurately detects the original scam video title and description (left). However, simple keyword obfuscations, highlited as red, evades the text based detector (right). At the top, representative video frames illustrate visual scam signals, such as gift card code generation and personal information requests, which remain undetected by the text model. Notably the video contains irrelevant background audio.\footnote{\url{https://www.youtube.com/watch?v=_NeG7clfZnw}}} 
\label{fig:text_limitattion}
\end{figure*}

%% file: tables/prompt_scam.tex
\begin{table*}[h]
\centering

\resizebox{\linewidth}{!}{
\begin{tabular}{p{0.18\linewidth} p{0.78\linewidth}}
\hline
\textbf{Description} & \textbf{Prompt Contents} \\ \hline

Task & The provided frames are extracted from a YouTube video. This video has been identified as a scam. Your task is to analyze the content and determine the most relevant reasons for this classification, using the following official YouTube Terms and Conditions violations. \\ \hline

Rules & 
1. Claims to commit a crime on behalf of the user, regardless of whether it actually does. \newline
2. Purports to provide an “unbounded” giveaway that offers unlimited free items without rules, limit or end. \newline
3. Promises viewers they'll see something but instead directs them off-site. \newline
4. Gets clicks, views, or traffic off YouTube by promising viewers that they'll make money fast. \newline
5. Sends audiences to sites that can spread malware, try to gather personal information or other sites that have a negative impact. \newline
6. Offers cash gifts, get rich quick schemes, or pyramid schemes. \newline
7. Impersonates an individual, company, or organization. \\ \hline

Instructions & 
1. Carefully examine the frames by paying close attention to all details in the content. \newline
2. Select one or more reasons from the list above that justify why this video qualifies as a scam. \newline
3. For each selected reason, provide a brief explanation grounded in the content. \\ \hline

Output Format & [Reason(s) in full text]: Brief explanation for each selected reason. \\ \hline

\end{tabular}
}
\caption{Prompt used for GPT assisted reasoning generation for scam videos.}
\label{tab:gpt_prompt_scam}
\end{table*}

%% file: tables/prompt_nonscam.tex
\begin{table*}[!ht]
\centering

\setlength{\tabcolsep}{5pt}       
\begin{tabular}{p{0.18\linewidth} p{0.75\linewidth}}
\hline
\textbf{Description} & \textbf{Prompt Contents} \\ \hline

Task & The provided frames are extracted from a YouTube video. This video has been identified as a non-scam. Your task is to analyze the content and briefly explain why this video appears legitimate. Consider the following official YouTube Terms and Conditions violations as indicators of scams. \\ \hline

Rules &
1. Claims to commit a crime on behalf of the user, regardless of whether it actually does.  \newline
2. Purports to provide an “unbounded” giveaway that offers unlimited free items without rules, limit, or end.  \newline
3. Promises viewers they'll see something but instead directs them off-site.  \newline
4. Gets clicks, views, or traffic off YouTube by promising fast money.  \newline
5. Sends audiences to sites that spread malware, gather personal information, or cause harm.  \newline
6. Offers cash gifts, get-rich-quick schemes, or pyramid schemes.  \newline
7. Impersonates an individual, company, or organization. \\ \hline

Instructions &
1. Carefully examine the frames by paying close attention to all details in the content.  \newline
2. Provide a brief summary of what the video is about, emphasizing its legitimate nature.  \newline
3. Avoid listing all seven points; instead, summarize why the video is a non-scam. \\ \hline

Output Format & A short explanation highlighting the legitimate purpose of the video, without enumerating every violation. \\ \hline

\end{tabular}
\caption{Prompt used for GPT-assisted reasoning generation for non-scam videos.}
\label{tab:gpt_prompt_nonscam}
\end{table*}

%% file: scam.bib
@misc{natlee2024blurgenerator,
  author ={Nat Lee},
  title        = {Blur-Generator},
  howpublished = {\url{https://github.com/NatLee/Blur-Generator}},
  year         = {2024},
  note         = {Accessed: January 2026},
}

@misc{yt-dlpgit,
  author   = {{yt-dlp}},
  title    = {yt-dlp},
  howpublished     = {\url{https://github.com/yt-dlp/yt-dlp}},
  note     = "Accessed: January 2026",
  year = {2025}
}

@misc{YouTube_search,
    author = {YouTube},
    title    = {YouTube search API},
    howpublished = {\url{https://developers.google.com/youtube/v3/docs/search/list}},
    note     = "Accessed: January 2026",
    year = {2025}
}

@misc{Google-translation-API,
    author = {Google},
    title ={Cloud Translation API},
    howpublished = {\url{https://cloud.google.com/translate/docs/reference/rest}},
    note     = "Accessed: January 2026",
    year = {2025}
}

@misc{NordVPN,
    author ={NordVPN},
    title    = {Is this link safe?},
    howpublished = {\url{https://nordvpn.com/link-checker}},
    note = "Accessed: January 2026",
    year = {2025}
}

@misc{VirusTotal,
    author = {VirusTotal},
    title    = {Analyse suspicious files, domains, IPs and URLs to detect malware and other breaches, automatically share them with the security community. },
    howpublished = {\url{https://www.virustotal.com/gui/home/url}},
    note     = "Accessed: January 2026",
    year = {2025}
}

@article{mcdonald2019reliability,
  title={Reliability and inter-rater reliability in qualitative research: Norms and guidelines for CSCW and HCI practice},
  author={McDonald, Nora and Schoenebeck, Sarita and Forte, Andrea},
  journal={Proceedings of the ACM on human-computer interaction},
  volume={3},
  number={CSCW},
  pages={1--23},
  year={2019},
  publisher={ACM New York, NY, USA}
}

@misc{MtrevisoKrippendorffalphaPython,
  author = {Thomas Grill},
  title = {Mtreviso/Krippendorff-Alpha: {{Python}} Implementation of {{Krippendorff}}'s Alpha That Supports Multilabel Data -- Inter-Rater Reliability},
  howpublished = {\url{https://github.com/mtreviso/krippendorff-alpha}},
  year = {2025}
}

@misc{demoji,
    author    = {Brad Solomon},
    title ={Accurately find or remove emojis from a blob of text using data},
    howpublished = {\url{https://pypi.org/project/demoji}},
    note     = "Accessed: January 2026",
    year = {2025}

}

@misc{videoblog,
    author ={Mark Polskii},
    title = {The Emergence of Video-Centric Social Networks: Trends and Impacts},
    howpublished = "\url{https://inappstory.com/blog/video-centric-social-media}",
    note = "Accessed: July 2025",
    year = {2025}
}

@misc{youtubestat,
    author = {GMI Research Team},
    title = {YOUTUBE statistics 2026 (Demographics, users by country and more)},
    howpublished = "\url{https://www.globalmediainsight.com/blog/youtube-users-statistics/}",
    note = "Accessed: July 2025",
    year = {2025}
}

@misc{gardner2018allennlpdeepsemanticnatural,
      title={AllenNLP: A Deep Semantic Natural Language Processing Platform}, 
      author={Matt Gardner and Joel Grus and Mark Neumann and Oyvind Tafjord and Pradeep Dasigi and Nelson Liu and Matthew Peters and Michael Schmitz and Luke Zettlemoyer},
      year={2018},
      eprint={1803.07640},
      archivePrefix={arXiv},
      primaryClass={cs.CL},
      url={https://arxiv.org/abs/1803.07640}, 
}

@misc{youtube_policy,
  author       = "{YouTube Help}",
  title        = "{YouTube Policies on Spam, deceptive practices, and scams policies}",
  howpublished = "\url{https://support.google.com/youtube/answer/2801973?hl=en}",
  note         = "Accessed: July 2025",
  year = {2025}
}

@article{bouma2021first,
  title={A first look at scams on youtube},
  author={Bouma-Sims, Elijah and Reaves, Brad},
  journal={arXiv preprint arXiv:2104.06515},
  year={2021}
}

@article{tripathi2022analyzing,
  title={Analyzing the uncharted territory of monetizing scam Videos on YouTube},
  author={Tripathi, Ashutosh and Ghosh, Mohona and Bharti, Kusum},
  journal={Social Network Analysis and Mining},
  volume={12},
  number={1},
  pages={119},
  year={2022},
  publisher={Springer}
}

@article{li2023towards,
  title={Towards understanding and characterizing the arbitrage bot scam in the wild},
  author={Li, Kai and Guan, Shixuan and Lee, Darren},
  journal={Proceedings of the ACM on Measurement and Analysis of Computing Systems},
  volume={7},
  number={3},
  pages={1--29},
  year={2023},
  publisher={ACM New York, NY, USA}
}

@misc{krippendorff2011computing,
  title={Computing Krippendorff's alpha-reliability},
  author={Krippendorff, Klaus},
  year={2011}
}

@article{gupta2019angel,
  title={Angel or Demon? Characterizing Variations Across Twitter Timeline of Technical Support Campaigners},
  author={Gupta, Srishti and Bhatia, Gurpreet Singh and Suri, Saksham and Kuchhal, Dhruv and Gupta, Payas and Ahamad, Mustaque and Gupta, Manish and Kumaraguru, Ponnurangam},
  journal={The Journal of Web Science},
  volume={6},
  year={2019}
}

@inproceedings{larson2018using,
  title={Using web-scale graph analytics to counter technical support scams},
  author={Larson, Jonathan and Tower, Bryan and Hadfield, Duane and Edge, Darren and White, Christopher},
  booktitle={2018 IEEE International Conference on Big Data (Big Data)},
  pages={3968--3971},
  year={2018},
}

@inproceedings{miramirkhani2016dial,
  title={Dial one for scam: Analyzing and detecting technical support scams},
  author={Miramirkhani, Najmeh and Starov, Oleksii and Nikiforakis, Nick},
  booktitle={22nd Annual Network and Distributed System Security Symposium (NDSS)},
  volume={16},
  year={2016}
}

@inproceedings{srinivasan2018exposing,
  title={Exposing search and advertisement abuse tactics and infrastructure of technical support scammers},
  author={Srinivasan, Bharat and Kountouras, Athanasios and Miramirkhani, Najmeh and Alam, Monjur and Nikiforakis, Nick and Antonakakis, Manos and Ahamad, Mustaque},
  booktitle={Proceedings of the 2018 World Wide Web Conference},
  pages={319--328},
  year={2018}
}

@inproceedings{kharraz2018surveylance,
  title={Surveylance: Automatically detecting online survey scams},
  author={Kharraz, Amin and Robertson, William and Kirda, Engin},
  booktitle={2018 IEEE Symposium on Security and Privacy (SP)},
  pages={70--86},
  year={2018},
}

@inproceedings{badawi2019game,
  title={The “game hack” scam},
  author={Badawi, Emad and Jourdan, Guy-Vincent and Bochmann, Gregor and Onut, Iosif-Viorel and Flood, Jason},
  booktitle={International Conference on Web Engineering},
  pages={280--295},
  year={2019},
  organization={Springer}
}

@article{suarez2019automatically,
  title={Automatically dismantling online dating fraud},
  author={Suarez-Tangil, Guillermo and Edwards, Matthew and Peersman, Claudia and Stringhini, Gianluca and Rashid, Awais and Whitty, Monica},
  journal={IEEE Transactions on Information Forensics and Security},
  volume={15},
  pages={1128--1137},
  year={2019},
  publisher={IEEE}
}

@inproceedings{al2020social,
  title={Social-guard: Detecting scammers in online dating},
  author={Al-Rousan, Suhaib and Abuhussein, Abdullah and Alsubaei, Faisal and Kahveci, Ozkan and Farra, Hazem and Shiva, Sajjan},
  booktitle={2020 IEEE International Conference on Electro Information Technology (EIT)},
  pages={416--422},
  year={2020},
}

@inproceedings{liu2024give,
  title={Give and take: An end-to-end investigation of giveaway scam conversion rates},
  author={Liu, Enze and Kappos, George and Mugnier, Eric and Invernizzi, Luca and Savage, Stefan and Tao, David and Thomas, Kurt and Voelker, Geoffrey M and Meiklejohn, Sarah},
  booktitle={Proceedings of the 2024 ACM on Internet Measurement Conference},
  pages={704--712},
  year={2024}
}

@inproceedings{li2023double,
  title={Double and nothing: Understanding and detecting cryptocurrency giveaway scams},
  author={Li, Xigao and Yepuri, Anurag and Nikiforakis, Nick},
  booktitle={Proceedings of the Network and Distributed System Security Symposium (NDSS)},
  year={2023}
}

@inproceedings{acharya2024conning,
  title={Conning the crypto conman: End-to-end analysis of cryptocurrency-based technical support scams},
  author={Acharya, Bhupendra and Saad, Muhammad and Cin{\`a}, Antonio Emanuele and Sch{\"o}nherr, Lea and Dai Nguyen, Hoang and Oest, Adam and Vadrevu, Phani and Holz, Thorsten},
  booktitle={2024 IEEE Symposium on Security and Privacy (SP)},
  pages={17--35},
  year={2024},
}

@inproceedings{agarwal2025hey,
  title={‘Hey mum, I dropped my phone down the toilet’: Investigating Hi Mum and Dad SMS Scams in the United Kingdom},
  author={Agarwal, Sharad and Harvey, Emma and Mariconti, Enrico and Suarez-Tangil, Guillermo and Vasek, Marie and others},
  booktitle={Usenix Security Symposium},
  year={2025}
}

@article{mishra2020smishing,
  title={Smishing Detector: A security model to detect smishing through SMS content analysis and URL behavior analysis},
  author={Mishra, Sandhya and Soni, Devpriya},
  journal={Future Generation Computer Systems},
  volume={108},
  pages={803--815},
  year={2020},
  publisher={Elsevier}
}

@inproceedings{kotzias2025ctrl+,
  title={Ctrl+ Alt+ Deceive: Quantifying User Exposure to Online Scams.},
  author={Kotzias, Platon and Pachilakis, Michalis and Aldana-Iuit, Javier and Caballero, Juan and S{\'a}nchez-Rola, Iskander and Bilge, Leyla},
  booktitle={NDSS},
  year={2025}
}

@inproceedings{kolupuri2025scams,
  title={Scams and frauds in the digital age: ML-based detection and prevention strategies},
  author={Kolupuri, Sai Venkata Jaswant and Paul, Ananya and Bhowmick, Rajat Subhra and Ganguli, Isha},
  booktitle={Proceedings of the 26th International Conference on Distributed Computing and Networking},
  pages={340--345},
  year={2025}
}

@inproceedings{vakilinia2022cryptocurrency,
  title={Cryptocurrency giveaway scam with youtube live stream},
  author={Vakilinia, Iman},
  booktitle={2022 IEEE 13th Annual Ubiquitous Computing, Electronics \& Mobile Communication Conference (UEMCON)},
  pages={0195--0200},
  year={2022},
}

@inproceedings{chu2022behind,
  title={Behind the tube: Exploitative monetization of content on {YouTube}},
  author={Chu, Andrew and Arunasalam, Arjun and Ozmen, Muslum Ozgur and Celik, Z Berkay},
  booktitle={31st USENIX Security Symposium (USENIX Security 22)},
  pages={2171--2188},
  year={2022}
}

@inproceedings{bouma2025kids,
  title={The Kids Are All Right: Investigating the Susceptibility of Teens and Adults to YouTube Giveaway Scams.},
  author={Bouma-Sims, Elijah Robert and Klucinec, Lily and Lanyon, Mandy and Downs, Julie and Cranor, Lorrie Faith},
  booktitle={NDSS},
  year={2025}
}

@article{hu2022lora,
  title={Lora: Low-rank adaptation of large language models.},
  author={Hu, Edward J and Shen, Yelong and Wallis, Phillip and Allen-Zhu, Zeyuan and Li, Yuanzhi and Wang, Shean and Wang, Lu and Chen, Weizhu and others},
  journal={ICLR},
  volume={1},
  number={2},
  pages={3},
  year={2022}
}

@misc{yang2024qwen2technicalreport,
      title={Qwen2 Technical Report}, 
      author={An Yang et al.},
      year={2024},
      eprint={2407.10671},
      archivePrefix={arXiv},
      primaryClass={cs.CL},
      url={https://arxiv.org/abs/2407.10671}, 
}

@inproceedings{zhai2023sigmoid,
  title={Sigmoid loss for language image pre-training},
  author={Zhai, Xiaohua and Mustafa, Basil and Kolesnikov, Alexander and Beyer, Lucas},
  booktitle={Proceedings of the IEEE/CVF international conference on computer vision},
  pages={11975--11986},
  year={2023}
}

@misc{radford2022robustspeechrecognitionlargescale,
      title={Robust Speech Recognition via Large-Scale Weak Supervision}, 
      author={Alec Radford and Jong Wook Kim and Tao Xu and Greg Brockman and Christine McLeavey and Ilya Sutskever},
      year={2022},
      eprint={2212.04356},
      archivePrefix={arXiv},
      primaryClass={eess.AS},
      url={https://arxiv.org/abs/2212.04356}, 
}

@article{zhang2019bertscore,
  title={Bertscore: Evaluating text generation with bert},
  author={Zhang, Tianyi and Kishore, Varsha and Wu, Felix and Weinberger, Kilian Q and Artzi, Yoav},
  journal={arXiv preprint arXiv:1904.09675},
  year={2019}
}

@article{shengyu2023instruction,
  title={Instruction tuning for large language models: A survey},
  author={Shengyu, Zhang and Linfeng, Dong and Xiaoya, Li and Sen, Zhang and Xiaofei, Sun and Shuhe, Wang and Jiwei, Li and Hu, Runyi and Tianwei, Zhang and Wu, Fei and others},
  journal={arXiv preprint arXiv:2308.10792},
  year={2023}
}

@inproceedings{alberto2015tubespam,
  title={Tubespam: Comment spam filtering on youtube},
  author={Alberto, T{\'u}lio C and Lochter, Johannes V and Almeida, Tiago A},
  booktitle={2015 IEEE 14th international conference on machine learning and applications (ICMLA)},
  pages={138--143},
  year={2015},
}

@inproceedings{chaudhary2013contextual,
  title={Contextual feature based one-class classifier approach for detecting video response spam on youtube},
  author={Chaudhary, Vidushi and Sureka, Ashish},
  booktitle={2013 Eleventh Annual Conference on Privacy, Security and Trust},
  pages={195--204},
  year={2013},
}

@inproceedings{zannettou2018good,
  title={The good, the bad and the bait: Detecting and characterizing clickbait on youtube},
  author={Zannettou, Savvas and Chatzis, Sotirios and Papadamou, Kostantinos and Sirivianos, Michael},
  booktitle={2018 IEEE Security and Privacy Workshops (SPW)},
  pages={63--69},
  year={2018},
}

@inproceedings{oak2025hello,
  title={" Hello, is this Anna?": Unpacking the Lifecycle of {Pig-Butchering} Scams},
  author={Oak, Rajvardhan and Shafiq, Zubair},
  booktitle={Twenty-First Symposium on Usable Privacy and Security (SOUPS 2025)},
  pages={1--18},
  year={2025}
}

@article{burton2024pig,
  title={Pig butchering in cybersecurity: A modern social engineering threat},
  author={Burton, Sharon L and Moore, Pamela D},
  journal={SocioEconomic Challenges},
  volume={8},
  number={3},
  pages={46},
  year={2024},
  publisher={Academic Research and Publishing UG}
}

@inproceedings{morris2020textattack,
  title={TextAttack: A Framework for Adversarial Attacks, Data Augmentation, and Adversarial Training in NLP},
  author={Morris, John and Lifland, Eli and Yoo, Jin Yong and Grigsby, Jake and Jin, Di and Qi, Yanjun},
  booktitle={Proceedings of the 2020 Conference on Empirical Methods in Natural Language Processing: System Demonstrations},
  pages={119--126},
  year={2020}
}

@inproceedings{ilavendhan2024optimizing,
  title={Optimizing YouTube Spam Detection with Ensemble Deep Learning Techniques},
  author={Ilavendhan, A and Janani, N and others},
  booktitle={2024 14th International Conference on Cloud Computing, Data Science \& Engineering (Confluence)},
  pages={625--630},
  year={2024},
}
